\documentstyle[art12,bezier,emlines2]{article}
\normalbaselines
\begin{document}
\voffset=-2cm

\begin{center}
{\Large {\bf On inert properties of particles in classical theory}}\\[12mm]
{\bf B.\ P.\ Kosyakov} \\[3mm]
{\it Russian Federal Nuclear Center -- All-Russia Research Institute of
Experimental Physics, Sarov, 607190 Russia}.

{E-mail address:\verb|kosyakov@vniief.ru|} \\[15mm]
\end{center}

\begin{abstract}
\noindent 
This is a critical review of inert properties of classical relativistic point 
objects.
The objects are classified as Galilean and non-Galilean. 
Three types of non-Galilean objects are considered: spinning, rigid, and 
dressed particles. 
In the absence of external forces, such
particles are capable of executing not only uniform motions along
straight lines but also Zitterbewegungs,  self-accelerations, 
self-decelerations, and uniformly accelerated motions. 
A free non-Galilean object possesses the four-velocity and the four-momentum 
which are in general not collinear, therefore, its inert properties are 
specified 
by two, rather than one, invariant quantities. It is shown that a spinning 
particle need not be a non-Galilean object. 
The necessity of a rigid mechanics for the construction
of a consistent classical electrodynamics in spacetimes of dimension
$D+1$ is justified for $D+1>4$. 
The problem of how much the form 
of fundamental laws of physics orders four dimensions of our world is
revised together with its solution suggested by Ehrenfest. The present
analysis made it apparent that the notion of the ``back-reaction''
does not reflect the heart of the matter in classical field
theories with point-like sources,  the notion of ``dressed'' particles
proves more appropriate.
\end{abstract}

\tableofcontents

\section{Introduction}
\label
{introduction} 
In modern textbooks on classical field theory (see, e.\ g., 
\cite{Landau}--\cite{Barut}) the concept of inertia of point-like
objects has received not too much attention. 
Some authors are concerned with a parameter of the appropriate dimension 
in the mechanical part of the Lagrangian identifying it with mass in the
Newtonian sense, familiar from the school physics, while another authors
derive relativistic concepts less formally, in an ``inductive'' way. 
Yet, whatever premises, the line of reasoning is
basically aimed at the indoctrination of the universal signi\-fi\-cance 
(both on classical and quantum levels) of
the quantity $M$ defined by the relation $p^{2}=M^{2}$.
This quantity is called mass, with no adjectives. 
Many people think of it as {\it the only} quantity specifying
inert properties of particles.

However, for a more rigorous treatment, particular emphasis should
be placed upon the context. 
For example, if we are dealing with a classical picture, 
the quantity $M$ alone is insufficient. 
Experts are well aware of this fact. 
However, they use to ``feel too shy'' to mention it in journals for the
general physical audience and textbooks.

Put very simply, the essence of the problem is this.
States of a relativistic point-like object may be characterized by the 
four-coordinate $x^{\mu}$ in Minkowski space and the four-momentum $p^{\mu}$. 
On the classical level, the four-velocity $v^{\mu}=dx^{\mu}/ds$, $s$ being the
proper time, is also well defined. From the vectors $p^{\mu}$ and $v^{\mu}$,
two invariants can be built: 
\begin{equation}
M^{2}=p^{2}  \label{M-def}
\end{equation}
and 
\begin{equation}
m=p\cdot v,  \label{m-def}
\end{equation}
while the invariant $v^{2}=1$ is dynamically trivial, it
manifests only the parametrization choice.
$M$ and $m$ are called,
respectively, the {\it mass} and the {\it rest mass}. 
For {\it Galilean particles}, these quantities are 
numerically equal.
However, {\it non-Galilean} particles are also tolerable in classical theory. 
In the absence of external forces, such particles
can execute not only uniform motion, but also trembling, self-accelerating,
self-decelerating, and hyperbolic motions. A free non-Galilean object
possesses the four-velocity $v^{\mu}$ and the four-momentum $p^{\mu}$ which
in general are nonparallel, and hence inert properties of this object are
characterized by two {\it different} quantities $M$ and $m$.

Three types of non-Galilean objects, {\it spinning}, {\it rigid}, and 
{\it dressed} particles, are a central preoccupation of this review. 
Schr\"{o}dinger was the first to speak about a trembly regime, the
visualization of solutions to the Dirac equation; since then this phenomenon 
bears the expressive German name ``Zitterbewegung''. 
More recently a classical realization of this
pheno\-me\-non, a helical world line \cite{Huang,Corben}, was found. 
An evolution mode for a free dressed charged particle with exponentially
growing acceleration, see, e.\ g., \cite{Landau}--\cite{Barut}, was
discovered (presumably by Lorentz) at the end of the 19th century.
Although such solutions to the equations of motion are believed to be
``unphysical'', their very existence changes the view on the Galilean
motion as the exclusive regime for particles subjected to the 
``self-interaction'',
actually for every real particle since any one possesses some charge (electric,
Yang--Mills', or gravitational), hence being ``self-interacting''. The
capability of a dressed colored particle for the self-decelerated motion in
the absence of external forces was pointed out in \cite{k1}--\cite{k8}. The
fact that rigid particles can execute Zitterbewegungs and runaways has also
not gone unnoticed, see, e.\ g., \cite{KosyakovNesterenko} and references
therein.

Our discussion is restricted to the classical context, that is, any pure 
quantum problem is ruled out, and we do not touch on issues in curved 
spacetimes of general relativity. 
Quantum theory may be sporadically mentioned to make more
prominent the classical character of the subject. 
The set of allowable world lines is taken to be composed of only
smooth timelike or lightlike future-directed world lines, containing no
spacelike curves or fragments of such curves (associated with superluminal
motions), past-directed timelike curves (interpreted as the world lines of
anti-particles), and piecewise smooth timelike curves made of adjacent
future-directed and past-directed fragments. We leave aside any modification
of the notion of mass related to extensions of spacetime symmetries, in
particular the Schouten--Haantjes idea that mass behaves as a scalar
density of the weight $-1/2$  under
conformal transformations, see, e.\ g., \cite{Schouten}.
We dwell on elementary objects while the problem of mass of composite
systems receives little attention. 
We adopt the retarded (rather than advanced or some else) boundary
condition on classical fields generated by particles.

Among models of spinning particles, we address only J.\ Frenkel's
model \cite{frenkel}, the first description of a classical particle
with spin, and the model with Grass\-man\-nian variables for spin
degrees of freedom proposed by F.\ A.\ Berezin and M.\ S.\ Marinov \cite
{Berezin}, and R.\ Casalbuoni \cite{Casalbuoni}, and elaborated by C.\ A.\
P.\ Galvao and C.\ Teitelboim \cite{GalvaoTeitelboim}. We discuss rigid 
particles with acceleration-dependent Lagrangians, even though models with 
higher derivatives may
be found in the literature (for a complete list of references see 
\cite{Pavsic}). We are  concerned with dressed charged
and colored particles in four-dimensional Minkowski space ${E}_{1,3}$,
but we do not cover another dressed particles in ${E}_{1,3}$, e.\ g.,
particles interacting with scalar or tensor fields \cite{Barut75}, dressed
particles with spin interacting with electromagnetic \cite{RoweRowe,Barut89}, 
scalar and tensor fields \cite{BarutCruz}, dressed particles in curved
manifolds \cite{BrehmedeWitt, BarutVillarroel}, and in flat spaces of other
dimensions, for example, in ${E}_{1,5}$ \cite{k9}.

The purpose of these limitations is twofold. First, they reveal the presence
of {\it several} quantities specifying inertia even in 
this restricted scope and separate the nonuniqueness problem
in the given context from that in the general case.
Second, a careful analysis of the notion of mass is requisite for other 
contexts.
It is not improbable
that this task may seem attractive to the reader of this paper.

The paper is organized as follows. Section 2 deals with the problem of mass
for Galilean particles together with relevant issues omitted in the
educational literature. It transpires that Newton's second law is tailored
for the relativistic mechanics being smoothly embedded into the
four-dimensional geometry. We show how the embedding is accomplished. Two
forms of the action of Galilean particles are considered. Their equivalence
for a finite particle mass, and admissibility of only one of them
for massless particles are established. Section 3 offers an account of inert
properties of the Frenkel particle. In the modern formulation, this model is
simple (at least in the absence of interactions) and instructive, but its
desctiption is scattered over research papers. Because of
this, the model is discussed carefully and in the form convenient for the
introductory learning. The fact that spinning particles do not necessarily
behave as non-Galilean objects is demonstrated by the example of the model
with Grassmannian variables in Sec.\ 4. Section 5 is devoted
to the problem of inert properties of rigid particles. Motivations for
studies of the higher derivative dynamics are given in Sec.\ 6. It is
shown that the construction of a consistent classical electrodynamics in
spacetimes of dimensions $D+1>4$ leads inevitably to the notion of rigid
particles. In this connection, we revise the problem of the
four-dimensionality of our world together with its solution suggested by P.\
Ehrenfest. In Sec.\ 7,  the notion of the ``dressed'' particle is shown to be 
more adequate for classical theories with point-like sources than the
notion of the ``self-interacting'' particle. The problem of mass is
illustrated by two comparatively simple examples of dressed particles. The
summary of the discussion and points in favor of it are in Sec.\ 8.

The paper is intended mainly for readers with the
basic knowledge of standard field theory. That is why major issues are
self-contained whenever possible\footnote{
Such a detailing may be justified by the fact that the analyzed problem
still went unnoticed not only in the special monographic literature (the
well-known book \cite{Jammer} including), but also in essays for the general
physical audience.}, and, hopefully, their understanding will not require to
consult original papers.

For the most part, we use standard notations. Gaussian units are adopted,
the speed of light $c$ and elementary quantum $\hbar$ are set to 1. 
The metric is $\eta_{\mu\nu}={\rm diag}\,(+,-,-,-)$. Repeated Greek
indices take the values 0, 1, 2, 3, while Latin indices run from 1 to 3. In
some instances an obvious coordinate free geometric symbolism is applied to
four-dimensional quantities, and three-dimensional vectors are denoted by
boldface letters. World lines are parametrized either by an affine
parameter $\lambda $ (derivative w.\ r.\ t.\ $\lambda $ is denoted by a
prime) or with the aid of the proper time $s$ (derivative w.\ r.\ t.\ $s$
is denoted by a dot). 
The special symbols $v^{\mu}$ and $a^{\mu}$ stand,
respectively, for the four-velocity ${\dot{x}}^{\mu}$ and four-acceleration $%
{\ddot{x}}^{\mu}$.

\section{Galilean particle}

\label{gp} 
Were we striving to embody the special relativity in
a single phrase, this intention is best expressed as follows: 
``{\it Spacetime of the physical world is described by pseudoeuclidean
four-dimensional geometry of the signature $+, -, -, -$}''. 
This means implicitly
that all dynamical laws are represented as geometric statements.

An adherent of the deductive method of the ``Course of Theoretical Physics''
by Landau and Lifshitz, who normally views the principle of least action as
Alpha and Omega of theoretical constructions, should meet such a geometric
encoding with sympathy. Indeed, given a geometry, we can determine geometric 
invariants, write down Lagrangians as all possible
invariant structures, and, varying the action, derive dynamical
equations. We, therefore, can formally whittle things down to setting the
geometry.

However, with closer inspection of the substantive aspect, it
would transpire that an important issue was overlooked. 
It is impossible to verify experi\-men\-tal\-ly the geometry by itself. 
The point, going back to Poincar\'e \cite{Poincare}, is what to be verified 
is just the totality of geometry and physical laws, or, in symbolic form, 
$\Gamma+\Phi$. 
Changing $\Gamma$, one can modify $\Phi$ in such a way that theoretical 
predictions of
phenomena are left intact. Therefore, it is insufficient to fix the
spacetime description (patterned after the pseudoeuclidean geometry), one
should also clarify the way the operationally well defined physical notions
(such as force, mass, energy and momentum) are incorporated into
the theory.

Let us make clear the status of the Newtonian dynamics. A
widespread misunder\-standing is that the second Newton's law in its {\it %
primordial} form 
\begin{equation}
\frac{d{\bf p}}{dt}={\bf f}  \label{newton-orig-p}
\end{equation}
does not work at velocities comparable with the speed of light, it must be
denounced in this domain, and the ``true law of the relativistic
mechanics'' 
\begin{equation}
\frac{d}{dt}\,m\gamma {\bf v}={\bf F},\quad \gamma =\frac{1}{\sqrt{1-{\bf v}%
^{2}}}  \label{planck-eq-p}
\end{equation}
derived by Planck in 1906 \cite{Planck6} must be accepted. In actual fact,
Eq.\ (\ref{newton-orig-p}) need neither be rejected, nor modified, it should
be only {\it smoothly embedded} it into the four-dimensional geometry of
Minkowski space, which automatically yields Eq.\ (\ref{planck-eq-p}).

\begin{figure}[htb]
\begin{center}
\unitlength=1mm
\special{em:linewidth 0.4pt}
\linethickness{0.4pt}
\begin{picture}(38.00,34.00)
\bezier{60}(20.00,5.00)(20.00,16.00)(22.00,20.00)
\bezier{44}(22.00,20.00)(24.00,27.00)(26.00,30.00)
\put(20.00,10.00){\vector(0,1){8.00}}
\put(23.00,24.00){\vector(1,3){3.33}}
\emline{23.00}{24.00}{1}{38.00}{28.00}{2}
\emline{23.00}{24.00}{3}{11.00}{21.00}{4}
\emline{11.00}{10.00}{5}{34.00}{10.00}{6}
\put(30.00,7.00){\makebox(0,0)[cc]{$\Sigma$}}
\put(35.00,24.00){\makebox(0,0)[cc]{$\Sigma$}}
\put(18.00,18.00){\makebox(0,0)[cc]{$v$}}
\put(24.00,34.00){\makebox(0,0)[cc]{$v$}}
\bezier{36}(26.00,30.00)(28.00,33.00)(27.00,38.00)
\end{picture}
\caption{The hyperplane $\Sigma$ perpendicular  to the world line}
\label
{hyperplane}
\end{center}
\end{figure}
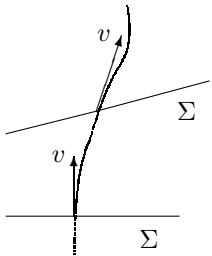

The idea of the embedding is rested on the fact that Eq.\ 
(\ref{newton-orig-p}) becomes an {\it asymptoti\-cal\-ly exact} law as 
${\bf v}\rightarrow 0$. 
This means that Eq.\ (\ref{newton-orig-p}) describes quite
correctly the dynamics in an {\it instantaneously accompanying} inertial
frame of reference where the velocity of the object is ${\bf v}=0$, or, in
the geometric language, the vector relation (\ref{newton-orig-p}) is exact
on the hyperplane $\Sigma $ perpendicular to the world line. Meanwhile the
hyperplane $\Sigma $ rotates together with the normal vector $v^{\mu }$ as
one travels along the world line, Figure \ref{hyperplane}. 
Thus the algorithm for construction of a global relativistic picture is to
jump in the instantaneously accompanying inertial frame, read and
execute the local dynamical prescription (\ref{newton-orig-p}), 
jump in the next instantaneously accompanying frame, etc. In other
words, for the embedding, we need an operator $\stackrel{\scriptstyle v}{%
\bot }$\thinspace\ that would permanently project vectors of Minkowski space
on the hyperplanes $\Sigma $ perpendicular to the world line. This operator is 
\begin{equation}
\stackrel{\scriptstyle v}{\bot }_{\hskip0.5mm\mu \nu }\,=\,{\eta }_{\mu \nu
}-\frac{v_{\mu }v_{\nu }}{v^{2}},  \label{projector-def}
\end{equation}
and the projection of any vector $X^{\mu }$ on the hyperplane $\Sigma $ is 
\begin{equation}
(\stackrel{\scriptstyle v}{\bot }X)^{\hskip0.3mm\mu }=X^{\mu }-\frac{X\cdot v%
}{v^{2}}\,v^{\mu }.  \label{projection-X}
\end{equation}
The formulas (\ref{projector-def}) and (\ref{projection-X}) are valid also
for arbitrarily parametrized world lines, one should only change $v$ by 
${x^{\prime}}$. 
As to the parametrization by the proper time, we are
dealing with even simpler formulas, because $v^{2}=1$.

Consider how the projector (\ref{projector-def}) embeds the one-parameter
family of the three-dimen\-sional equations (\ref{newton-orig-p}) in four
dimensions. 
In the instantaneously accompanying frame,
the time axis $t$ is aligned with the tangent to the world line at the given
instant, thus $dt$ coincides with $ds$ (this follows formally from the
relation $ds=\gamma ^{-1}dt$ where $\gamma \rightarrow 1$ as 
${\bf v}\rightarrow 0$), and the differentiation w.\ r.\ t.\ $t$ can be replaced by
the differentiation w.\ r.\ t.\ $s$. 
From the {\it Newtonian three-force} ${\bf f}$, one can uniquely regain 
the {\it Minkowski four-force} $f^{\mu }$. 
Indeed, components of $f^{\mu }$ in an arbitrary frame of
reference originate (through the Lorentz boost) from components of this
vector in the rest frame where, {\it by definition}, they are 
\begin{equation}
f^{\mu }=(0,\,{\bf f}).  \label{f-rest-syst}
\end{equation}
Define the four-momentum $p^{\mu }$ in such a way that the
derivative of its spatial components w.\ r.\ t.\ $s$ coincide with components 
of the three-vector $d{\bf p}/dt$ in Eq.\ (\ref{newton-orig-p}) in the 
accom\-pa\-nying frame of reference. Then the required embedding is
\begin{equation}
\stackrel{\scriptstyle v}{\bot }({\dot{p}}-f)=0.  \label{newton-symbl}
\end{equation}

We notice that the projector (\ref{projector-def}) is defined only on
timelike tangent vectors and makes no sense for isotropic tangent vectors,
thus solutions to Eq.\ (\ref{newton-symbl}) describe only smooth {\it 
timelike} world lines.

Mechanical objects of different types reveal different dependences of $%
p^{\mu }$ on kine\-ma\-ti\-cal variables. The simplest possibility provides
an {\it elementary Galilean object}. This is a point-like object. Its states
in the nonrelativistic limit are specified by the three-coordinate of its
location ${\bf x}$ and its three-momentum 
\begin{equation}
{\bf p}=m{\bf v}.  \label{p=mv-3}
\end{equation}
Such objects are usually called {\it particles}, with no adjectives. We will
follow this tradition, though one should bear in mind that we cover not the
total set of point-like mechanical objects but only its Galilean subset.

The N\thinspace e\thinspace w\thinspace t\thinspace o\thinspace n\thinspace
i\thinspace a\thinspace n\thinspace\ m\thinspace a\thinspace s\thinspace
s\thinspace\ $m$ is a fundamental characteristic of the particle
\footnote{The particle may have another quantities specifying its 
individuality, for example, couplings with different fields. 
But, unlike mass specifying the particle ``intrinsically'', these quantities
characterize it relative to other objects.}, it remains constant no matter
how great the influence on the particle (that is, under every possible force 
${\bf f}$): 
\begin{equation}
\frac{d}{dt}\,m=0.  \label{dm=0}
\end{equation}
The ``elementary'' character of the object will be understood as the
impossibility of its splitting which is formally controlled by the condition
(\ref{dm=0})\footnote{%
The condition (\ref{dm=0}) is consistent with the assumption that $f^{\mu }$
in the rest frame takes the form (\ref{f-rest-syst}). If we put $f^{\mu }=(k,%
{\bf f})$, rather than (\ref{f-rest-syst}), then, apart from (\ref
{newton-symbl}), the equation ${\dot{m}}=k$ would arise.}.

With the expression (\ref{p=mv-3}) for ${\bf p}$, Eq.\ (\ref{newton-orig-p})
reduces to 
\begin{equation}
m{\bf a}={\bf f}.  \label{ma=f-3}
\end{equation}
For ${\bf f}=0$, Eq.\ (\ref{ma=f-3}) has a unique solution ${\bf v}$ =
const. 
Thus free particles evolve in the {\it Galilean regime}.

From (\ref{p=mv-3}), it is clear that the particle four-momentum $p^{\mu}$
is 
\begin{equation}
p^{\mu}=mv^\mu.  \label{p=mv-4}
\end{equation}
In view of the relation $v\cdot a=0$, the projector 
$\stackrel{\scriptstyle v}{\bot}$ in Eq.\ (\ref{newton-symbl}) acts as a unite 
operator, and this
equation is reduced to 
\begin{equation}
ma^\mu=f^\mu.  \label{ma=f-4}
\end{equation}

Since the Minkowski four-force is orthogonal to the four-velocity, $f\cdot
v=0$, components of $f^\mu$ in an arbitrary Lorentz frame of reference are
not independent, they are related by $f^0=f^i v^i$. It is convenient to
separate $\gamma$ as an overall factor of $f^\mu$: 
\begin{equation}
f^\mu=\gamma\,({\bf F}\cdot{\bf v},\,{\bf F}).  \label{f-mu-decomp}
\end{equation}
Then ${\bf F}$ is found to be the three-force in the {\it Planck sense}
because the spatial component of Eq.\ (\ref{ma=f-4}) acquires the form (\ref
{planck-eq-p}). As to the time component, 
\begin{equation}
\frac{d}{dt}\,m\gamma={\bf F}\cdot{\bf v},  \label{work}
\end{equation}
it may be interpreted as the equation of variation of energy 
${\cal E}=m\gamma$ due to the work performed by the force ${\bf F}$ in a
unite time.

So, the replacement of (\ref{newton-orig-p}) by (\ref{planck-eq-p}) does not
imply that the Newtonian dynamics, as such, has been subjected to a revision
or modification, it demonstrates only that Newton's second law has been
smoothly embedded into the four-dimensional pseudoeuclidean geometry.
(Fixing such a kind of geometry, we thus have maximally ``loaded'' $\Gamma $
and left a minimum of the ``load'' for $\Phi $ -- it is just the virtue of 
the pseudoeuclidean model of
spacetime.)

The m\,a\,s\,s\, $M$\, and\, r\,e\,s\,t\, m\,a\,s\,s\, $m$ of a particle are
defined by Eqs.\ (\ref{M-def}) and (\ref{m-def}). 
With the expression for $p^{\mu}$, Eq.\ (\ref{p=mv-4}), $M$ and $m$ are 
identical to one
another and the Newtonian mass (denoted also by $m$). The latter
remains constant not only when the particle is free, but also under the
action of any force (this is called for by the convention
of the elementary character of the particle).

The equality $M=m$ is crucial for  the equivalence of mass
and rest energy. 
Note in this connection that the concept of inertia is not replenished with 
a  ``relativistic'' content. 
All the conceptual novelty of relativistic dynamics of Galilean
particles, as opposed to the Newtonian dynamics, amounts to the mere
geometrical fact that energy and momentum are temporal and
spatial components of the timelike vector $p^{\mu }$ of length $M$, and,
therefore, the particle energy ${\cal E}$ is reckoned from $M$, rather than 0.

We now turn to the Lagrangian and Hamiltonian formulations. Let us see the
way the projective structure of Eq.\ (\ref{newton-symbl}) is related to
symmetries of the theory. The action ${\cal A}$ depends on the world line
configuration, rather than the parametrization, and hence ${\cal A}$ remains
unchanged under the reparametrization transformations 
\begin{equation}
\lambda=\lambda(\xi),\quad x^\mu(\lambda)=x^\mu\bigl(\lambda(\xi)\bigr)
\label{reparam}
\end{equation}
where $\lambda(\xi)$ is an arbitrary continuous monotonic function of $\xi$.
We represent the reparametrizations in the
infinitesimal form: 
\begin{equation}
\delta\lambda=\epsilon,\quad \delta x^\mu=\epsilon\,{x^{\prime}}^\mu,
\label{reparam-inf}
\end{equation}
$\epsilon=\epsilon(\xi)$ is an arbitrary infinitesimal continuous positive
function of $\xi$. From the invariance of ${\cal A}$ under the
transformations (\ref{reparam-inf}) follows the identity\footnote{%
Note that the identity (\ref{Noether-reparam}) is the simplest illustration
of the second Noether theorem \cite{Noether} in the case of the
infinite-dimensional group of transformations (\ref{reparam}) leaving the
action ${\cal A}$ invariant.} 
\begin{equation}
\frac{\delta {\cal A}}{\delta x^\mu}\,{x^{\prime}}^\mu=0
\label{Noether-reparam}
\end{equation}
which just implies that the Eulerian $\delta {\cal A}/\delta x^\mu$ involves
the projective operator $\stackrel{\scriptstyle v}{\bot}$\,. Thus, given a
reparametrization invariant action, this provides the embedding of the
Newtonian dynamics into spacelike hyperplanes $\Sigma$ (which embody the
ordinary three-space in instanta\-neous\-ly accompanying frames of
reference).

The reparametrizations (\ref{reparam}) is a kind of local gauge
transformations; their analog in general relativity is general coordinate
transformations $x^\mu=x^\mu(y)$, the so called diffeomorphisms of the
pseudo-Riemannian space. Extending the dynamical frame\-work, we may, along
with $p^\mu$ of the form (\ref{p=mv-4}), consider any conceivable dependence
of the momentum on kinematical variables, yet the reparametrization
invariance requirement, or, equivalently, the presence of the projective
structure remains therewith indisputable.

It might be well to point out that the projector $\stackrel{\scriptstyle v}{%
\bot}$ acts as a unite operator {\it solely} for $p^\mu=mv^\mu$. Thus, it
would be erroneously to think, as is, alas, the case, that Eq.\ (\ref{ma=f-4}%
) is the equation of the relativistic dynamics in a broad sense. Such a role
is assigned to Eq.\ (\ref{newton-symbl}). It is the equation that describes
the evolution of any structureless mechanical object of finite mass.

The action for a relativistic Galilean particle proposed by Planck \cite
{Planck6}, 
\begin{equation}
{\cal A}=-\mu\int dt\,\sqrt{1-{\bf v}^2}  \label{Planck-action}
\end{equation}
is readily rewritten in the reparametrization invariant form: 
\begin{equation}
{\cal A}= -\mu\int d\lambda\,\sqrt{{x^{\prime}}\cdot{x^{\prime}}}.
\label{Planck-action-repar}
\end{equation}
The variation of (\ref{Planck-action}) w.\ r.\ t.\ ${\bf v}$ gives the
canonical three-momentum 
\begin{equation}
{\bf p}=\mu\gamma{\bf v},  \label{p-Planck}
\end{equation}
while the variation of (\ref{Planck-action-repar}) w.\ r.\ t.\ ${x^{\prime}}%
^\alpha$ gives the canonical four-momentum 
\begin{equation}
p^{\alpha}= -\frac{\delta{\cal A}}{\delta {x^{\prime}}_{\alpha}} =\frac{\mu\,%
{x^{\prime}}^\alpha}{\sqrt{{x^{\prime}}\cdot{x^{\prime}}}}\,.
\label{p-mu-Planck}
\end{equation}
The expression (\ref{p-mu-Planck}) is coincident with the expression (\ref
{p=mv-4}) when $\mu=m $. Therefore, the {\it formal parameter} $\mu$ should
be identified with the quantities $m$ and $M$, and also the Newtonian mass.
We will find in the following that such quantities specifying non-Galilean
object are all distinct.

One further reparametrization invariant action for a Galilean particle
proposed by L.\ Brink, P.\ Di Vecchia and P.\ Howe is 
\begin{equation}
{\cal A}=-\frac{1}{2}\,\int d\lambda \,\biggl(\frac{{x^{\prime }}^{2}}{\eta }%
+\eta \,\mu ^{2}\biggr).  \label{Brink-action}
\end{equation}
Here, $\eta (\lambda )$ is an auxiliary variable; the reader aware of
elements of general relativity may interpret it as the square root of the
determinant of the one-component metric world line tensor $\sqrt{\det
g_{\lambda \lambda }}=\sqrt{g_{\lambda \lambda }}$ since its transformation
law under the reparametrizations (\ref{reparam}) is 
\[
\eta (\lambda )=
\frac{d\xi }{d\lambda }\,\eta \bigl(\lambda (\xi )\bigr). 
\]
In the literature, the quantity $\eta ^{-1}$ is referred to as ``Einbein''
or ``monad''.

We now can define the m\,a\,s\,s\,l\,e\,s\,s\, Galilean particle as an
object for which $\mu=0$. The action (\ref{Brink-action}) for such a
particle is nonzero.

The variation of the action (\ref{Brink-action}) w.\ r.\ t.\ $\eta$ gives
the constraint equation 
\begin{equation}
-\eta^{-2}\,{x^{\prime}}^2 +\mu^2=0  \label{eq-eta}
\end{equation}
from which in the case $\mu\ne 0$ we find 
\begin{equation}
\eta^{-1}=\mu\,({{x^{\prime}}\cdot{x^{\prime}}})^{-1/2}.  \label{solution}
\end{equation}
When $\lambda$ is realized as the ``laboratory time'' $t$, we have $%
\eta^{-1}=\mu/\sqrt{1-{\bf v}^2}$, that is, the monad is identical to energy 
${\cal E}$ of the particle\footnote{%
In the subnuclear physics, this fact provokes occasionally the temptation to
construe the quantities $\mu$ and $\eta^{-1}$ as, respectively, the {\it %
current} and {\it constituent} masses of quarks \cite{Simonov}. The reason
for this is that light, $u$ and $d$, quarks confined in hadrons behave as
ultra-relativistic objects for which ${\cal E}\gg\mu$, hence a plausible
explanation of great difference of values of the current and constituent
masses of such quarks.}, while, for $\lambda=s$, we have $\eta^{-1}=\mu$.

Substituting (\ref{solution}) in (\ref{Brink-action}), we revert to the
Planck action (\ref{Planck-action-repar}).

Notice, for $\mu\ne 0$, the actions (\ref{Planck-action-repar}) and (\ref
{Brink-action}) are equivalent on the quantum level as well. This can
readily be verify by means of the Feynman path integral. Indeed, when
employing the action (\ref{Brink-action}), the path integral involves an
additional integration over $\eta(\lambda)$ that can be worked out through
the use of the well-known result for the one-dimensional integration: 
\[
\int^\infty_0 \frac{d\eta}{\sqrt{\eta}}\,\exp\biggl(-\frac{A}{\eta}- B\eta%
\biggr)=\sqrt{\frac{\pi}{B}}\,\exp(-2\sqrt{AB}), \quad A>0,\, B>0. 
\]

From the action (\ref{Brink-action}), we obtain the expression for the
canonical momentum: 
\begin{equation}
p^\alpha=\eta^{-1}{x^{\prime}}^\alpha.  \label{p-mu-Brink}
\end{equation}
With it, the constraint (\ref{eq-eta}) is written in the form 
\begin{equation}
-p^2+\mu^{2}=0  \label{p-sqr=mu-sqr}
\end{equation}
which is identical to (\ref{M-def}) when $M=\mu$.

The Hamiltonian corresponding to the action (\ref{Brink-action}) is 
\begin{equation}
H={p}\cdot{x^{\prime}}+L={\frac12}\,\eta\,(p^2-\mu^2).  \label{H-Galileo}
\end{equation}
It is notable that $H=0$ on the constraint (\ref{p-sqr=mu-sqr}). {\it Zero
Hamiltonians} are generally inherent in reparametrization invariant models.
[The Hamiltonian corresponding to the action (\ref{Planck-action-repar}) is
identically zero.] The action (\ref{Brink-action}) is then representable in
the form quite clear from the canonical formalism viewpoint: 
\begin{equation}
{\cal A}= \int d\lambda\,\bigl(-{p}\cdot{x^{\prime}}+H)= \int d\lambda\,%
\bigl[-{p}\cdot{x^{\prime}}+ {\frac12}\,\eta\,(p^2-\mu^2)\bigr]
\label{Brink-action-p-q}
\end{equation}
where $\eta$ may be interpreted as the Lagrangian multiplier of the problem
with the constraint (\ref{p-sqr=mu-sqr}).

For $\mu=0$, the actions (\ref{Planck-action-repar}) and (\ref{Brink-action}%
) are {\it nonequivalent}. To describe a massless particle one should
evidently proceed from the action (\ref{Brink-action}). (Notice, there is a
number of ``desert islands'' in this region, for one, the problem of the
motion of a massless particle under the action of some simple force.)

\section{Pure gyroscope}

\label{gyroscope} 
As the first example of non-Galilean objects, we look at Frenkel's spinning 
particle \cite{frenkel} (another name is pure gyroscope) following largely to  
\cite{Rafanelli}. 
For alternative treatments  see, e.\ g., \cite{Corben, Barut, HansonRegge}. 
We consider a free particle, that is, the interaction with any field
(in particular gravitational) is negligible. 
Because spacetime is homogeneous and isotropic, the
four-momentum $p^{\mu}$ and angular momentum tensor 
\begin{equation}
J_{\mu\nu}=x_\mu p_\nu-x_\nu p_\mu+\sigma_{\mu\nu},  \label{J-mu-nu}
\end{equation}
$\sigma_{\mu\nu}$ being a skew-symmetric real-valued spin angular momentum
tensor, are conserved quantities. 
Write down explicitly the conservation laws: 
\begin{equation}
{\dot p}^{\mu}=0,  \label{dot-p}
\end{equation}
\begin{equation}
{\dot J}_{\mu\nu}=0.  \label{dot-M}
\end{equation}
It is beyond reason to augment them by the addition of the spin conservation
law ${\dot\sigma_{\mu\nu}}=0$ since no extra symmetry is suggested. 
By (\ref{J-mu-nu})--(\ref{dot-M}), we have 
\begin{equation}
{\dot\sigma}_{\mu\nu}=p_\mu v_\nu-p_\nu v_\mu,  \label{dotsigma}
\end{equation}
thus the four-velocity and four-momentum of the spinning particle are in
general not collinear.

The pure gyroscope is defined as such a particle that 
\begin{equation}
\sigma _{\mu \nu }v^{\nu }=0.  \label{sigma-v}
\end{equation}

To understand the geometric meaning of this constraint, write a general
2-form $\sigma$ as a combination of exterior products of 1-forms (covectors) 
${\vec v},\,{\vec e}_1,\,{\vec e}_2$, ${\vec e}_3$ which span a moving
basis: 
\begin{equation}
\sigma=\sum_i\,{\cal D}_i\,{\vec v}\wedge{\vec e}_i+\sum_{i<j} {\cal K}%
_{ij}\,{\vec e}_i\wedge{\vec e}_j.  \label{sigma-decomp-gen}
\end{equation}
Let this basis be orthonormal at any intstant: 
\begin{equation}
{\vec v}^{ 2}=1,\quad {\vec v}\cdot {\vec e}_i=0,\quad {\vec e}_i\cdot {\vec
e}_j=-\delta_{ij}.  \label{ortog}
\end{equation}
Insertion of (\ref{sigma-decomp-gen}) in (\ref{sigma-v}) gives
\footnote{If the spinning 
particle is electrically charged, the Lagrangian involves the so called Pauli 
term proportional to $\sigma_{\mu\nu}F^{\mu\nu}$. 
In view of (\ref{sigma-decomp-gen}) and (\ref{ortog}), this term 
takes the form ${\bf d}\cdot{\bf E}+ {\bf m}\cdot{\bf B}$
where the electric dipole moment $d_i$ is proportional to ${\cal D}_i$, and
the magnetic dipole moment $m_i$ is proportional to ${\scriptstyle\frac12}
\epsilon_{ijk}{\cal K}_{jk}$. 
From (\ref{D-zero}) follows that the pure
gyroscope corresponds to a spinning particle with a magnetic dipole moment, 
but with no electric dipole moment. 
Due to precession of ${\bf m}$ around magnetic lines of force, 
this object is referred to as the ``gyroscope''.} 
\begin{equation}
{\cal D}_i=0.  \label{D-zero}
\end{equation}
Now, only three terms in (\ref{sigma-decomp-gen}) are left: 
\[
\sigma={\cal K}\,({\vec e}_1\wedge {\vec e}_2+\alpha\, {\vec e}_1\wedge {%
\vec e}_3+\beta\, {\vec e}_2\wedge {\vec e}_3). 
\]
Using the bilinearity and skew-symmetry of exterior products, the expression
in the parenthesis can be identically transformed to 
\[
{\vec e}_1\wedge ({\vec e}_2+\alpha {\vec e}_3)-\beta\, {\vec e}_3\wedge {%
\vec e}_2=({\vec e}_1-\beta {\vec e}_3) \wedge({\vec e}_2+\alpha {\vec e}%
_3). 
\]
Introducing two new base 1-forms ${\vec f}_1={\vec e}_1-\beta {\vec e}_3$
and ${\vec f}_2={\vec e}_2+\alpha {\vec e}_3$, we arrive at 
\begin{equation}
\sigma={\cal K}\,{\vec f}_1\wedge {\vec f}_2.  \label{sigma-red}
\end{equation}
Equation (\ref{sigma-red}) does not alter when the 1-form ${\vec f}_1$ is
substituted by the 1-form 
\[
{\vec g}_1={\vec f}_1-\frac{{\vec f}_1\cdot {\vec f}_2}{{\vec f}_2^{ 2}}\,{%
\vec f}_2 
\]
orthogonal to ${\vec f}_2$. 
Furthermore, we may normalize ${\vec g}_1$ and ${\vec f}_2$, and attribute 
their magni\-tu\-des to ${\cal K}$. We single out
the Planck constant $\hbar$ as an overall normalization in ${\cal K}$ to
yield 
\[
{\cal K}\,\sqrt{{\vec g}_1^{ 2} {\vec f}_2^{ 2}}={\cal S}{\hbar}. 
\]
We now return to the initial notations of
the base 1-forms, viz., ${\vec v},\,{\vec e}_1,\,{\vec e}_2$, ${\vec e}_3$,
rather than ${\vec v},\,{\vec g}_1,\,{\vec f}_2$, ${\vec e}_3$, stand
hereafter for the resulted basis.
This is quite legitimate, since only  ${\vec v}$ is fixed (being cotangent 
to the world line) while the
remainder of the basis is determined by the orthonormalization condition.
The net result (recall that $\hbar=1$) is 
\begin{equation}
\sigma={\cal S}\,{\vec e}_1\wedge {\vec e}_2  \label{sigma-canon}
\end{equation}
where ${\cal S}$ is the spin magnitude in the rest frame in which 
$v^{\mu}=(1,\,0,\,0,\,0)$, $e_1^{\mu}=(0,\,1,\,0,\,0)$, and $e_2^{\mu}=
(0,\,0,\,1,\,0)$. Equation (\ref{sigma-canon}) shows that ${\cal S}$ can be
defined in an invariant way: 
\begin{equation}
\sigma_{\mu\nu}\sigma^{\mu\nu}=2{\cal S}^2.  \label{sigma-sqr}
\end{equation}
One further useful relation derivable from (\ref{sigma-canon}) is 
\begin{equation}
\sigma_{\lambda\mu}\sigma^{\mu\nu}\sigma_{\lambda\rho}=- {\cal S}%
^2\sigma_{\lambda\rho}.  \label{sigma-cube}
\end{equation}

Let us return to the equation of the spin evolution (\ref{dotsigma}). 
From (\ref{sigma-v}), we obtain 
\[
\sigma_{\mu\nu}{\frac{d}{ds}}\,\sigma^{\mu\nu}={\frac12}\, {\frac{d}{ds}}%
\,\sigma_{\mu\nu}\sigma^{\mu\nu}=0. 
\]
In view of (\ref{sigma-sqr}), it follows that ${\cal S}$ is $s$-independent.

The gyroscope mass $M$ and rest mass $m$ are defined by (\ref{M-def}) and 
(\ref{m-def}). 
Since we consider a free object, we assume that $p^{\mu}$ is timelike
future-directed vector. Thus, $M^{2}>0$, $p^0>0$, and $m>0$. 
By (\ref{dot-p}), $M$ is constant in time. We will see later that $m=$ const as well.

We define the quantity 
\begin{equation}
\zeta^\mu= \sigma^{\mu\nu} p_\nu.  \label{zeta-def}
\end{equation}
From (\ref{zeta-def}) and (\ref{sigma-v}) it follows 
\begin{equation}
{\zeta}\cdot {p}=0, \quad {\zeta}\cdot {v}=0.  \label{zeta-v}
\end{equation}
With (\ref{sigma-canon}), we find 
\[
{\zeta}^\mu= {\cal S}\,[{e}^\mu_1\,({e}_2\cdot {p})-{e}^\mu_1\,({e}_2\cdot {p%
})\,]. 
\]
It is evident now that ${\zeta}^\mu$ is a spacelike vector, 
\begin{equation}
{\zeta}^{ 2}<0.  \label{zeta-sqr}
\end{equation}
Equation (\ref{dotsigma}) can be recast in the form 
\begin{equation}
{\dot\zeta}^\mu=-M^2 v^\mu+m p^{\mu}.  \label{dot-zeta}
\end{equation}
Contraction with ${\zeta}_{\mu}$ yields ${\zeta}^{2}={\rm const}$. This
means that only the direction of $\zeta^\mu$ varies in time, not the
magnitude.

Differentiation of (\ref{sigma-cube}) w.\ r.\ t.\ $s$, contraction with $%
v^\rho$, and making use of (\ref{dotsigma}) leads to 
\begin{equation}
m v^\lambda=p^{\lambda}+\frac{1}{{\cal S}^2}\,\sigma^{\lambda\mu}
\zeta_{\mu},  \label{p-v-rel}
\end{equation}
and further contraction with $p_\lambda$ results in 
\begin{equation}
m^2=M^2-\frac{{\zeta}^{2}}{{\cal S}^2}.  \label{m-M-rel}
\end{equation}
We see that $m$ is a constant of motion because such are quantities in the
right hand side of (\ref{m-M-rel}). Combining (\ref{m-M-rel}) and (\ref
{zeta-sqr}), we conclude that 
\begin{equation}
m^2>M^2.  \label{m>M}
\end{equation}

Why $M\ne m$? 
They differ since $v^{\mu}$ and $p^{\mu}$ are not collinear.
To see this, differentiate (\ref{p-v-rel}) and take into account (\ref{dot-p}), (\ref{dotsigma}), (\ref
{zeta-v}), (\ref{dot-zeta}), (\ref{sigma-v}), and (\ref{zeta-def}).
The result is
\[
{\cal S}^2{\dot v}^{\mu}=\zeta^{\mu}. 
\]
Further differentiation leads to 
\begin{equation}
{\cal S}^2{\ddot v}^{\mu}+M^2v^{\mu}=m p^{\mu}.  \label{eq-motion}
\end{equation}
One easily observes the similarity of (\ref{eq-motion}) with the
equation of harmonic oscillator under the action of an external constant
force. Thus a solution is 
\begin{equation}
v^{\mu}(s)=\frac{m}{M^2}\,p^{\mu}-\alpha^{\mu}\sin \omega
s+\beta^{\mu}\cos\omega s  \label{v-solution}
\end{equation}
where $\alpha\cdot p=\beta\cdot p=\alpha\cdot\beta=0$,\, $\alpha^2=\beta^2$.
Integration provides the world line: 
\begin{equation}
x^{\mu}( s)=\frac{m}{M^2}\,p^{\mu} s+\frac{\alpha^{\mu}} {\omega}\,\cos
\omega s+\frac{\beta^{\mu}}{\omega}\,\sin \omega s.  \label{x-solution}
\end{equation}
This is a helical world line, a realization of the Zitterbewegung. The
rotation with the frequency $\omega=M/{\cal S}$ occurs on the plane spanned
by two vectors $\alpha^{\mu}$ and $\beta^{\mu}$, perpendicular to the vector 
$p^{\mu}$. The amplitude of the rotation $\sqrt{\alpha^2}=\sqrt{\beta^2}$
being equal to the projection of the vector $p^\mu$ onto the plane spanned
by two vectors $e^\mu_1$ and $e^\mu_2$ is arbitrary while the period of the
rotation $T=2\pi{\cal S}/M$ is of order of the Compton wave length of
the particle, $1/M$.

If we assume that $p^2<0$, then (\ref{x-solution}) is replaced by 
\begin{equation}
x^{\mu}( s)=-\frac{m}{{\cal M}^2}\,p^{\mu} s+\frac{\alpha^{\mu}}{\Omega}\, {%
\cosh}\,\Omega s+\frac{\beta^{\mu}}{\Omega}\,{\sinh}\,\Omega s
\label{x-hyperb}
\end{equation}
where ${\cal M}^2=-p^2$, $\Omega={\cal M}/{\cal S}$, and $\alpha^{\mu}$ and 
$\beta^{\mu}$ meet the condition $\alpha^2=-\beta^2$. 
This solution, describing a motion across the plane spanned by two vectors 
$p^\mu$ and $\alpha^\mu$, shows an enhancement of velocity. 
The solution (\ref{x-solution}) corresponds to a compactly supported motion, 
while the solution (\ref{x-hyperb}) corresponds to the motion in a noncompact 
region. 
If the momentum space is limited by the condition $p^2\ge 0$, 
the configuration space contains the Zitterbewegung (\ref{x-solution}) 
but is free of the runaway (\ref{x-hyperb}).

Averaging (\ref{v-solution}) over $s$ gives 
\begin{equation}
<{v}^{\mu}>\,=\frac{m}{M^2}\,p^{\mu}.  \label{v-ave}
\end{equation}
Let us trace the motion of the point with the coordinate 
\begin{equation}
y^\mu= x^\mu+\frac{1}{M^2}\,\zeta^\mu.  \label{y-def}
\end{equation}
With (\ref{dot-zeta}) and (\ref{v-ave}), we have 
\begin{equation}
{\dot y}^\mu=\frac{m}{M^2}\,p^{\mu}=\,<{v}^\mu>.  \label{dot-y}
\end{equation}
The point with the coordinate $y^\mu$ draws a straight world line with the
guiding vector $p^{\mu}$. 
This point is interpreted as the center of mass. 
The conserved four-momentum $p^{\mu}$ must be assigned
to an imagined carrier which is located at
the center of mass and moves along the averaged world line.

The availability of two masses gives rise to two spins. 
Indeed, one may define spin as the internal angular momentum
related to either {\it kinematical rest frame} where ${\bf v}=0$, 
i.~e., $v^{\mu}=(1,\,0,\,0,\,0)$, or {\it dynamical rest frame} where 
${\bf p}=0$,
i.~e., $p^{\mu}=M\,(1,\,0,\,0,\,0)$. So far we discussed the former
possibility. In order to turn to the latter, we should, as is clear from (%
\ref{y-def}) and (\ref{dot-y}), to use the notion of the center of mass. We
express $x_\mu$ through $y_\mu$ and substitute the result in (\ref{J-mu-nu})
to yield 
\begin{equation}
J_{\mu\nu}=y_\mu p_\nu-y_\nu p_\mu+\Xi_{\mu\nu}  \label{J-via-s}
\end{equation}
where the tensor 
\begin{equation}
\Xi_{\mu\nu}=\sigma_{\mu\nu}-(\zeta_\mu p_\nu-\zeta_\nu p_\mu)/M^2
\label{s-def}
\end{equation}
plays now the same role as did $\sigma_{\mu\nu}$. In fact, the relation 
\begin{equation}
\Xi_{\mu\nu}p^\nu=0  \label{Xi-p}
\end{equation}
is an analog of the constraint (\ref{sigma-v}), and hence relations
analogous to (\ref{sigma-canon})--(\ref{sigma-cube}) take place, in
particular 
\begin{equation}
\Xi_{\mu\nu}\Xi^{\mu\nu}=2{S}^2,  \label{Xi-sqr}
\end{equation}
with $S=$ const, and 
\begin{equation}
{\dot \Xi}_{\mu\nu}=0  \label{dot-Xi}
\end{equation}
substitutes (\ref{dotsigma}). Taking the square of both sides of (\ref{s-def}%
), in view of (\ref{m-M-rel}), we find 
\begin{equation}
M^2{S}^2=m^2{\cal S}^2.  \label{M-s-rel}
\end{equation}
It is clear that the difference between ${S}$ and ${\cal S}$ is due to the
difference between $m$ and $M$.

All these results could be derived in a more regular way starting from the
Hamilto\-ni\-an 
\begin{equation}
H=\frac{1}{2}\,\eta \,\biggl(p^{2}-\mu ^{2}-\frac{{\zeta }^{2}}{{\cal S}^{2}}%
\biggr)  \label{H-gyro}
\end{equation}
and the canonical Poisson brackets 
\[
\{x_{\mu },x_{\nu }\}=\{p_{\mu },p_{\nu }\}=\{x_{\lambda },\sigma _{\mu \nu
}\}=\{p_{\lambda },\sigma _{\mu \nu }\}=0,\quad \{x_{\mu },p_{\nu
}\}=\eta_{\mu \nu }, 
\]
\begin{equation}
\{\sigma _{\mu \nu },\sigma _{\rho \sigma }\}=\sigma _{\mu \rho }\eta_{\nu
\sigma }+\sigma _{\nu \sigma }\eta_{\mu \rho }-\sigma _{\mu \sigma
}\eta_{\nu \rho }-\sigma _{\nu \rho }\eta_{\mu \sigma },
\label{Poisson-braket}
\end{equation}
rather than from ``heuristic'' equations (\ref{dot-p})--(\ref{dotsigma}).
The parametrization of the world line should be chosen such that the monad $%
\eta ^{-1}$ be fixed as $\eta ^{-1}=\mu $, and the parameter $\mu $ be
identified with $m$. Note that the expression in the parenthesis is the
constraint (\ref{m-M-rel}) whereby $H=0$. The Hamiltonian (\ref{H-gyro})
differs from that of Galilean particles (\ref{H-Galileo}) by the presence of
the last term generating the evolution of spin degrees of freedom.

Thus the inertia of a pure gyroscope is specified by two invariants, $M$ and 
$m$. In the absence of interactions, they are constant, and $m>M$ for all
time. This poses the dilemma: Which quantity of these two is measured by
experimenter? If only one of them is recorded in all cases, say, $m$, then
what is the reason for the prohibition from registration of another?
Alternatively, if the result of the measurement is equipment-dependent, what
is the peculiarity of the device that records, for example, only $M$?

While on the subject of a m\,a\,s\,s\,l\,e\,s\,s\, gyroscope, we encounter
new troubles. What should be a criterion of the masslessness: $M=0$ or $m=0$%
? If the masslessness is $M=0$, then the role of the positive invariant
conservative quantity $m$ is obscure. If, on the other hand, the
masslessness is $m=0$, then $p^2<0$, i.\ e., dynamically, the object behaves
as a {\it tachyon} (which, though, by no means suggests jumping through the
light barrier!).

Upon quantization, only a single of these two quantities, $M$ or $m$, may
survive. Which of them? Should the classical particle emerging in the limit $%
\hbar\to 0$ be massless (from some viewpoint) if the initial quantum
particle is massless?

Finally, one further dilemma is to decide between $M$ and $m$ in the
presence of gravitation; turning to the principle of {\it equivalence of
inert and gravitational masses}, one of these two quantities, $M$ or $m$,
should be set equal to the \,g\,r\,a\,v\,i\,t\,a\,t\,i\,o\,n\,a\,l\, mass $%
M_{{\rm g}}$. Which of them?

Now, following the Ortega y Gasset lessons \cite{Ortega}, we should reveal
honesty and tell the truth: We are dealing with a ``rebellion of the
masses''.

Another model of a classical spinning particle with $c$-number spinor
variables for the description of spin degrees of freedom was suggested by
A.\ O.\ Barut and N.\ Zanghi \cite{BarutZanghi}. In the absence of external
forces, such a particle behaves in a non-Galilean way, in many respects
similar to the pure gyroscope. All the above problems related to the
inequality $M\neq m$ remain here. We do not pause on this model since its
analysis would contribute little new to the present topic.

\section{Model with Grassmannian variables}

\label{grassmann} The issue of inert properties of a particle with spin
degrees of freedom described by real-valued Grassmannian variables $%
\theta^\mu$ and $\theta_5$ is another thing altogether. A refined version of
this model \cite{GalvaoTeitelboim} is specified by the reparametrization
invariant action 
\begin{equation}
{\cal A}=\int_{\lambda_1}^{\lambda_2} d\lambda\,\bigl[-p\cdot{x^{\prime}}+{%
\frac {\eta}{2}}\bigl(p^2-\mu^2\bigr)-{\frac{i}{2}}(\theta^{\prime}\cdot%
\theta+ {\theta^{\prime}}_5\theta_5)+i\chi(\theta\cdot p+\mu\theta_5) \bigr] %
-{\frac{i}{2}}[\theta(1)\cdot\theta(2)+\theta_5(1)\theta_5(2)]
\label{W-Grassmann}
\end{equation}
and the boundary variation conditions 
\begin{equation}
\delta x^\mu(1)=\delta x^\mu(2)=0,\quad
\delta\theta^\mu(1)+\delta\theta^\mu(2)=0, \quad
\delta\theta_5(1)+\delta\theta_5(2)=0.  \label{bound-Grassmann}
\end{equation}
The Grassmannian variable $\chi(\lambda)$ plays the role of a Lagrange
multiplier of the constraint.

From (\ref{W-Grassmann}) and (\ref{bound-Grassmann}), one derives four
dynamical equations 
\begin{equation}
{\dot{p}}^{\mu }=0,  \label{p'=0}
\end{equation}
\begin{equation}
-{\dot{x}}^{\mu }+\eta {p}^{\mu }+i\chi \theta ^{\mu }=0,
\label{p-x'-Grassmann}
\end{equation}
\begin{equation}
-{\dot{\theta}}^{\mu }+\chi {p}^{\mu }=0,  \label{p-theta-Grassmann}
\end{equation}
\begin{equation}
-{\dot{\theta}}_{5}+\chi \,\mu =0  \label{theta-5-chi-mu}
\end{equation}
(the proper time is chosen to be the parameter of evolution: $\lambda =s$)
and two constraints 
\begin{equation}
p^{2}-\mu ^{2}=0,  \label{p-sqr-mu-sqr}
\end{equation}
\begin{equation}
{\theta }\cdot p+\mu \,{\theta }_{5}=0.  \label{p-theta-constr}
\end{equation}

A first glance, the dependence between the momentum and velocity, (\ref
{p-x'-Grassmann}), is a direct analog of the dependence (\ref{p-v-rel})
responsible for the non-Galilean behavior of the pure gyroscope. But this
resemblance is deceptive. Indeed, since $\theta^0\theta^0=\theta^1\theta^1=
\theta^2\theta^2=\theta^3\theta^3=0$, it follows from (\ref{p-x'-Grassmann})
that 
\begin{equation}
({\dot x}^0-\eta {p}^0)^2=({\dot x}^1-\eta {p}^1)^2= ({\dot x}^2-\eta {p}%
^2)^2=({\dot x}^3-\eta {p}^3)^2=0.  \label{p-x'-sqr}
\end{equation}
We see that ${p}^\mu$ is parallel to ${\dot x}^\mu$. But ${\dot x}^\mu$ is a
timelike vector directed to the future, and hence, in view of (\ref
{p-sqr-mu-sqr}), $\eta^{-1}=\mu$. Equation (\ref{p-x'-Grassmann}) is
satisfied only for $\chi=0$. From (\ref{p-theta-Grassmann}) and (\ref
{theta-5-chi-mu}) follows that the Grassmannian variables do not vary in
time, ${\dot\theta}^\mu=0$, ${\dot\theta}_5=0$, while (\ref{p'=0}) and (\ref
{p-x'-Grassmann}) imply ${\dot x}^\mu=$ const. Thus the spin and
configuration variables evolve independently. The behavior of a free object
with the Grassmannian variables proves to be strictly Galilean. The object
is characterized by a single mass, $\mu=m=M$

It is obvious that the object capable of solely Galilean regime of evolution
will not identified with a non-Galilean object. Thus the model of spinning
particles with real-valued Grassmannian variables is not equivalent to the
model of spinning particles with spinor $c$-number variables \cite
{BarutZanghi}, contrary to the wrong assertion of Ref.\ \cite{BarutPavsic}.

\section{Rigid particle}

\label{rigid-particle} A point-like particle with the behavior governed by a
Lagrangian dependent on higher derivatives is our next example of
non-Galilean objects. Such a particle is called {\it rigid}. The velocity
and momentum of the rigid particle are in general nonparallel, it can
execute Zitterbewegung and runaway regimes. Thus the mass $M$ and rest mass $%
m$ of rigid particles are different quantities. Their dissimilarity is even
greater than that of the pure gyroscope: $M$ turns out to be a conserved
quantity while $m$ varies in time. Yet, it is not worth while to run ahead,
it would be better to discuss the subject in succession.

Recall that, by our convention, allowable world lines are only timelike
smooth curves. It would be sufficient for present purposes to consider a
reparametrization invariant action dependent on velocities and
accelerations, 
\begin{equation}
{\cal A}=\int_{\lambda_1}^{\lambda_2}
d\lambda\,L(x^{\prime},x^{\prime\prime}).  \label{W-rigid}
\end{equation}
It immediately follows that the Lagrangian may be written as 
\begin{equation}
L(x^{\prime},x^{\prime\prime})=\gamma^{-1}\Phi(k),  \label{L-sample}
\end{equation}
\begin{equation}
\gamma^{-1}=\sqrt{\,x^{\prime}\cdot x^{\prime}\,},  \label{gamma-general-def}
\end{equation}
where $\Phi(k)$ is an arbitrary function of the world line curvature $k$. It
is well known that the curvature squared is equal and of opposite sign to
the four-acceleration squared. We recall also that, in the general case of a
curve with an arbitrary parametrization $x^\mu(\lambda)$, the
four-acceleration $a^\mu$ is calculated from the formula: 
\begin{equation}
a^\mu=\gamma\,\frac{d}{d\lambda}\biggl(\gamma\, \frac{dx^\mu}{d\lambda} %
\biggr).  \label{a-mu-def}
\end{equation}

The Hamiltonian formalism of the rigid dynamics was originally developed in
a fundamental M.\ V.\ Ostrogradskii memoir back in 1850 \cite{Ostrogradskii}%
. We will need only some findings of this formalism \cite{Nesterenko} (the
derivation of them is left to the reader as a useful exercise). The
infinitesimal variation of the action can be represented in the form 
\begin{equation}
\delta {\cal A}=\int_{\lambda_1}^{\lambda_2}d\lambda\,\biggl[ \frac{\partial
L}{\partial x_\mu}-\frac{d}{d\lambda}\biggl( \frac{\partial L}{\partial {%
x^{\prime}}_\mu}\biggr)+\frac{d^2}{d\lambda^2}\biggl( \frac{\partial L} {%
\partial {x^{\prime\prime}}_\mu}\biggr)\biggr]\,{\bar\delta} x_\mu+(H\,
\delta\lambda-p\cdot\delta x- \pi\cdot\delta {x^{\prime}})
\vert_{\lambda_1}^{\lambda_2}  \label{delta-W}
\end{equation}
where the symbol ${\bar\delta}$ stands for the form variation of the world
line, ${\bar\delta} x_\mu={\delta} x_\mu-{x^{\prime}}_\mu\,{\delta}\lambda$, 
\begin{equation}
p^{\mu}= -\frac{\partial L}{\partial {x^{\prime}}_\mu}+\frac{d}{d\lambda}%
\biggl( \frac{\partial L}{\partial {x^{\prime\prime}}_\mu}\biggr),
\label{p-mu-rigid-def}
\end{equation}
\begin{equation}
\pi^\mu = -\frac{\partial L}{\partial {x^{\prime\prime}}_\mu},
\label{pi-mu-def}
\end{equation}
\begin{equation}
H = p\cdot {x^{\prime}}+\pi\cdot{x^{\prime\prime}}+L.  \label{H-rigid-def}
\end{equation}

As far as the Lagrangian $L$ is invariant under the four-coordinate
translations 
\begin{equation}
x_\mu\to x_\mu+c_\mu,  \label{x-transl}
\end{equation}
one infers (in line with the first Noether theorem) from (\ref{delta-W})
that $p^{\mu}$ is a constant of motion. On the other hand, the Lagrangian $L$
defies invariance under the four-velocity translations 
\begin{equation}
{x^{\prime}}_\mu\to {x^{\prime}}_\mu+d_\mu,  \label{v-transl}
\end{equation}
since this would conflict with the reparametrization invariance which is
assured by the presence in (\ref{L-sample}) of the quantity $\gamma$
non-invariant under the transformation (\ref{v-transl}), hence the canonical
momentum $\pi^\mu$ is not conserved. Besides, using formulas (\ref
{gamma-general-def}), (\ref{a-mu-def}) and (\ref{p-mu-rigid-def})--(\ref
{H-rigid-def}), one may check that the Hamiltonian $H=0$ for any Lagrangians
of the form (\ref{L-sample}); this is a consequence of the reparametrization
invariance of the action (\ref{W-rigid}). Thus $\pi^\mu$ and $H$ are
unusable for the determination of inert properties of a rigid particle, from
here on they will be of no interest.

Let the evolution parameter $\lambda $ be the proper time $s$. The
Lagrangian (\ref{L-sample}) yields the equation of motion 
\begin{equation}
(\stackrel{\scriptstyle v}{\bot }{\dot{p}})^{\hskip0.3mm\mu }=0.
\label{dot-p-rig}
\end{equation}
This equation shows plainly that, in the absence of external forces, the
canonical momentum $p^{\mu }$ is a conserved quantity. Thus the invariants (%
\ref{M-def}) and (\ref{m-def}) built from $p^{\mu }$ may characterize the
inertia of a rigid particle.

The problem of integration of the equation of motion (\ref{dot-p-rig}) in
the generic case of arbitrary smooth functions $\Phi(k)$ was investigated in 
\cite{NesterenkoFeoliScarpetta} where the interested reader is referred to
for detail. We will discuss here only a particular case 
\begin{equation}
\Phi(k)=-\mu+\nu k^2,  \label{Phi-sample}
\end{equation}
where $\mu$ and $\nu$ are arbitrary real parameters. We choose $\mu>0$
since, for $\nu=0$, one regains the Planck Lagrangian $L=-\mu\sqrt{%
\,x^{\prime}\cdot x^{\prime}\,}$ where $\mu$ is taken to be the rest mass $m$
of a Galilean particle. For the Lagrangian (\ref{Phi-sample}), one derives 
\begin{equation}
p^{\mu}=\mu v^\mu+\nu\,(2{\dot a}^\mu+3a^2 v^\mu).  \label{rig-momentum}
\end{equation}

Let the particle be moving along $z$-axis. Then $v^{\mu }$ may be
represented as follows 
\begin{equation}
v^{\mu }=({\cosh}\alpha,\,0,\,0,\,{\sinh }\alpha ).  \label{1D-motion}
\end{equation}
Differentiation gives higher derivative expressions, specifically, 
\[
a^{\mu}={\dot{\alpha}}\,({\sinh}\alpha,\,0,\,0,\,{\cosh}\alpha) 
\]
which implies $a^{2}=-{\dot{\alpha}}^{2}$. Equation (\ref{dot-p-rig})
reduces to 
\[
\mu {\dot{\alpha}}+\nu \,(2{\ddot{\alpha}}-{\dot{\alpha}}^{3})=0. 
\]
Denoting ${\dot{\alpha}}=q$ and ${\mu/\nu}=q_\ast^2$, rewrite it in the form 
\begin{equation}
{\ddot q}+{\frac12}\,q_\ast^2\,q-{\frac12}\,q^3=0.  \label{q-eq}
\end{equation}
The first integral of this equation is 
\begin{equation}
{\frac12}\,{\dot q}^{2}+U(q)={E},  \label{1st-int-q}
\end{equation}
\begin{equation}
U(q)=-{\frac18}\,(q^2-q_\ast^2)^2,  \label{U(q)-df}
\end{equation}
${E}$ is an arbitrary integration constant.

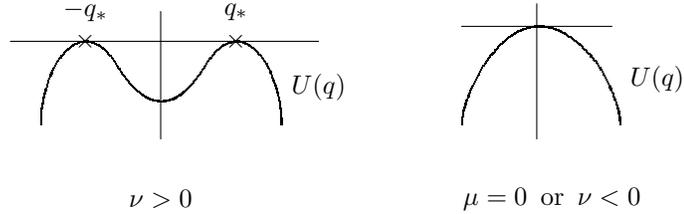
\begin{figure}[htb]
\begin{center}
\unitlength=1.00mm
\special{em:linewidth 0.4pt}
\linethickness{0.4pt}
\begin{picture}(95.00,30.00)
\emline{10.00}{23.00}{1}{51.00}{23.00}{2}
\emline{30.00}{10.00}{3}{30.00}{27.00}{4}
\bezier{92}(24.00,20.00)(30.00,10.00)(36.00,20.00)
\bezier{52}(36.00,20.00)(40.00,26.00)(44.00,20.00)
\bezier{52}(24.00,20.00)(20.00,26.00)(16.00,20.00)
\bezier{32}(16.00,20.00)(14.00,16.00)(14.00,12.00)
\bezier{36}(44.00,20.00)(46.00,16.00)(46.00,12.00)
\put(40.00,23.00){\makebox(0,0)[cc]{$\times$}}
\put(20.00,23.00){\makebox(0,0)[cc]{$\times$}}
\put(20.00,27.00){\makebox(0,0)[cc]{$-q_\ast$}}
\put(40.00,27.00){\makebox(0,0)[cc]{$q_\ast$}}
\put(51.00,17.00){\makebox(0,0)[cc]{$U(q)$}}
\emline{70.00}{25.00}{5}{90.00}{25.00}{6}
\emline{80.00}{10.00}{7}{80.00}{28.00}{8}
\bezier{96}(73.00,20.00)(80.00,30.00)(88.00,20.00)
\put(96.00,18.00){\makebox(0,0)[cc]{$U(q)$}}
\put(82.00,2.00){\makebox(0,0)[cc]{$\mu=0$\,\,\,{\rm or}\,\,\,$\nu<0$}}
\put(30.00,2.00){\makebox(0,0)[cc]{$\nu>0$}}
\bezier{36}(73.00,20.00)(70.00,15.00)(70.00,12.00)
\bezier{36}(88.00,20.00)(91.00,15.00)(91.00,12.00)
\end{picture}
\caption{The potential $U(q)$ of the fictitious particle}
\label
{potential}
\end{center}
\end{figure}

Equations (\ref{q-eq}) and (\ref{1st-int-q}) may be viewed as the equations
of motion of some fictitious particle of the unite mass in the potential
field $U(q)$. For $\nu>0$, the potential $U(q)$ has the shape schematically
depicted in Figure \ref{potential}, the left plot. For $-q_\ast^2/8<E<0$,
the motion of the fictitious particle is compactly supported, and falls in
the range $-q_\ast<q<q_\ast$. Thus, at not-too-large initial acceleration, $%
|a^2|<\mu /\nu$, the rigid particle executes a Zitterbewegung. For $E>0$, or 
$E<-q_\ast^2/8$, the fictitious particle executes an infinite motion. In
other words, if the initial acceleration of the rigid particle exceeds the
critical value $({\mu/\nu})^{1/2}$, a runaway regime is certainly realized.
For $E=0$, the fictitious particle rests on either of two tops of the
potential hill, that is, when $|a^2|=\mu/\nu$, the motion of the rigid
particle proves to be uniformly accelerated. However, this regime is
unstable, any small disturbance switches it to the runaway regime. For $%
E=-q_\ast^2/8$, the fictitious particle rests on the bottom of the potential
pit, which corresponds to the Galilean regime of the rigid particle, $a^2=0$%
. It is clear also that, for $\mu=0$, or $\nu<0$, the rigid particle is
capable of only a runaway regime, see the right plot in Figure \ref
{potential}. The Galilean regime of the rigid particle with such features is
unstable, any small disturbance switches it to a runaway regime. Thus the
instances with $\mu=0$ and $\nu<0 $ are of no physical interest.

With the aid of the Ansatz (\ref{1D-motion}), (\ref{rig-momentum}) can be
transformed to 
\[
p^\mu=(\mu-\nu{\dot\alpha}^2)({\cosh}\alpha,\,0,\,0,\,{\sinh}\alpha)+2\nu {%
\ddot\alpha}({\sinh}\alpha,\,0,\,0,\,{\cosh}\alpha). 
\]
It follows 
\[
p^2=(\mu-\nu{\dot\alpha}^2)^2-4\nu^2{\ddot\alpha}^2. 
\]
The comparison with (\ref{1st-int-q}) and (\ref{U(q)-df}) shows that 
$p^2=-8\nu^2E$. Thus the condition $p^2<0$ is tantamount to the condition 
$E>0 $ which is sufficient for the motion of the fictitious
particle to be infinite, or, what is the same, sufficient for the rigid
particle to be in a runaway motion.

Thus, if it is granted that the rigid particle is moving along a straight
line, and the four-momentum space is limited by the condition $p^2\ge 0$,
the only non-Galilean regime, the Zitterbewegung, may occur.

The non-Galilean motions are feasible not only on straight lines but also on
planes. 
Two regimes are realized here, the Zitterbewegung and
Zitterbewegungs with amplitudes enhanced in time. (If the Lagrangian depends
on velocities and accelerations, but independent of higher derivatives, the
dimension of the subspace $d$ where the Zitter\-be\-we\-gung occurs is 
no more than $d=2$ \cite{KosyakovNesterenko}.)

Expression (\ref{rig-momentum}) can be rewritten in a geometrically 
illuminating form: 
\begin{equation}
p^{\mu}=(\mu+\nu a^{2})\,v^{\mu}+ 2\nu\,(\stackrel{\scriptstyle v}{\bot}\!{%
\dot{a}})^{\hskip0.3mm\mu}  \label{rig-mom}
\end{equation}
where 
\[
(\stackrel{\scriptstyle v}{\bot}\!{\dot{a}})^{\hskip0.3mm\mu}= {\dot{a}}%
^{\mu}+a^{2}v^{\mu}. 
\]
It follows 
\begin{equation}
M^{2}=p^{2}=(\mu+\nu a^{2})^2+4\nu^2(\stackrel{\scriptstyle v}{\bot}\! {\dot{%
a}})^2,  \label{M-rig}
\end{equation}
\begin{equation}
m=p\cdot v=\mu +\nu a^{2}.  \label{m-rig}
\end{equation}

Thus both $M$ and $m$ reveal nontrivial dependences on kinematic
variables. 
Never\-theless, $M$ is constant.
As for $m$, it varies in time both in Zitterbewegung and runaway regimes. 
It is time-independent only for uniformly accelerated motions.
However, $p^{\mu}=0$ for such motions. 
This is clear from (\ref{rig-mom}) because the condition of relativistic 
uniformly accelerated motion is \cite{Rohrlich} 
\[
(\stackrel{\scriptstyle v}{\bot}\!{\dot a})^{\hskip0.3mm\mu}=0. 
\]

Thus $M$ is more fundamental than $m$ for the rigid particle. 
The parameters $\mu $ and $\nu $ in the Lagrangian (\ref{Phi-sample}) should 
be taken positive, even though they have no direct physical meaning. 
If rigid particles are realized in nature, their inert properties are most 
likely specified by $M$; just $M$ is expected to be be measured
experimentally. 
In quantum picture, the inertia of rigid
particles is represented by the sole quantity $M$. 
Moreover, just $M$ is natural to identify with the gravitational 
charge of the rigid particle $M_{{\rm g}}$ (provided that $M_{{\rm g}}$ is 
treated as a conserved quantity; with the advent of the idea of the black hole
evaporation \cite{Hawking}, the constancy of $M_{{\rm g}}$ became not so
evident, however, the consensus on this subject still remain to be found \cite
{abc}.).

Strange as it may seem, the problem of the mass of the rigid particle 
is more simple than that of the pure gyroscope.

\section{Why rigid particles?}
\label
{ng-why-rigid} 
Spinning and rigid particles are extravagant objects. 
While extravagances pertaining to spin are ``{\it Dei gratia}'', as the 
saying goes, phenomenological justi\-fi\-ca\-ti\-ons of peculiarities of 
the rigid dynamics still remain unknown. Is there a pure 
{\it theoretical} reason for recourse to the idea of higher derivatives?
Newton imagined no such reason. What changed in the past three hundred
years? What is the present-day role of the rigid dynamics? The following
claims are quite common in the literature: The rigid particle is a toy model
of rigid strings which in turn serve as a tool for effective description of
phase transitions in quantum chromodynamics \cite{Polyakov86,Kleinert};
properties of rigid particles are related to properties of hypothetical
anyons \cite{Kholodenko}; the rigidity is a useful concept in the polymer
chain physics \cite{KholodenkoViglis}, etc. 
It may well be that such arguments appear rather technical than fundamental.

To my mind, the dynamics with higher derivatives acquires its {\it raison
d'\^etre} in connection with the problem of the {\it consistency} of local
field theories in spaces of arbitrary dimensions. For example, let us extend 
the four-dimensional classical electrodynamics of
charged point-like particles to higher dimensions. 
Assume that the action is 
\begin{equation}
{\cal A}=-\sum_{i=1}^N\,\mu_i\int ds_i\,\sqrt{\,v_i\cdot v_i}-\sum_{i=1}^N\,
e_i \int dx_i^\mu A_\mu(x_i)-{\frac{1}{4\,\Omega_{D-1}}}\,\int
d^{D+1}\!x\,F_{\mu\nu}\,F^{\mu\nu}  \label{W-em}
\end{equation}
where $\Omega_{D-1}$ is the area of a $D-1$-dimensional sphere of unite
radius, and the field strength $F_{\mu\nu}$ is expressed through the
potential $A_{\mu}$ in the usual fashion: $F_{\mu\nu}=\partial_\mu
A_\nu-\partial_\nu A_\mu$. Given the action (\ref{W-em}), is it possible to
build such a classical theory where
all ultraviolet divergences are removed 
by some regularization-renormalization procedure?

Variation of the action w.\ r.\ t.\ $A_{\mu }$ gives the $D+1$-dimensional 
Maxwell equations 
\begin{equation}
\partial _{\mu }F^{\mu \nu }=\Omega _{D-1}j^{\nu },  \label{maxw}
\end{equation}
\begin{equation}
j^{\mu }(x)=\sum_{i=1}^{N}e_{i}\int_{-\infty }^{\infty }ds_{i}\,v_{i}^{\mu
}(s_{i})\,\delta ^{D+1}\biggl(x-x_{i}(s_{i})\biggr).  \label{j-mu}
\end{equation}
The retarded solutions to these equations for the motion of the charged 
particles along arbitrary timelike world lines $x_{i}^{\mu}(s_{i})$ are well 
known, see, e.\ g., \cite{k9}. We restrict our
consideration to the simplest case of a single
charge moving along a straight world line. 
Then the field is specified by a potential $\varphi ({\bf x})$, and Maxwell's 
equations (\ref{maxw})--(\ref{j-mu}) are reduced to the Poisson equation 
\begin{equation}
\Delta \varphi ({\bf x})=-\Omega _{D-1}\,\rho ({\bf x}),  
\label
{Poisson}
\end{equation}
\begin{equation}
\rho ({\bf x})=e\,\delta ^{D}({\bf x}).  
\label
{rho}
\end{equation}
The solution to (\ref{Poisson})--(\ref{rho}) is 
\begin{equation}
\varphi ({\bf x})=e\cases{\vert\,{\bf x}\vert^{2-D}, & $D\ne 2$,\cr
\log\vert\,{\bf x}\vert, & $D=2$.\cr}  \label{varphi}
\end{equation}
The electrostatic energy of the rest particle with the $\delta $-shaped
charge distribution (\ref{rho}), or,  the\thinspace\ s\thinspace
e\thinspace l\thinspace f-e\thinspace n\thinspace e\thinspace r\thinspace
g\thinspace y,\, is 
\begin{equation}
\delta m={\frac{1}{2}}\int d^{D}\!{\bf x}\,\rho({\bf x})\,\varphi({\bf x})=
\lim_{\epsilon\rightarrow 0}\,{\frac{1}{2}}\,e\,\varphi({\bf \epsilon}).
\label{delta m}
\end{equation}
By (\ref{varphi}), the self-energy $\delta m$ 
diverges linearly for $D=3$, while the divergence is cubic for $D=5$. 
These divergences
are due to the singular behavior of the fields at short distances from the
source, or, what is the same, slow decrease of the
Fourier-transforms of the fields at high frequencies, hence the name
``ultraviolet divergences''.

The standard approach to removal of these divergences is the infinite
renormalization of parameters appearing in the Lagrangian. 
Specifically,
we ascribe to the\, b\,a\,r\,e\, m\,a\,s\,s\, $\mu$ such a dependence on the
regularization parameter $\epsilon$ as to render the sum 
\begin{equation}
m=\lim_{\epsilon\to 0}\,\biggl(\mu(\epsilon)+\delta m(\epsilon)\biggr)
\label{ren-m}
\end{equation}
finite and positive. Then the r\,e\,n\,o\,r\,m\,a\,l\,i\,z\,e\,d\,
m\,a\,s\,s\, $m$ is maintained to be the rest mass of the particle. For
small $\epsilon$, the self-energy $\delta m$ becomes large positive, thus
the bare mass $\mu$ is large negative. However, this
is not a particular problem since $\mu$ and $\delta m$ come to view only at
intermediate stages and disappear once the passage to the limit (\ref
{ren-m}) is performed. 
They are not observable quantities. 
This status is assigned only to the renormalized mass $m$.

The renormalizability is a {\it necessary condition for consistency} of
local field theories \cite{Cao, SchweberR, k01}. 
Since processes of creation and annihilation of particles are missing 
from the classical picture
\footnote{Indeed, the set of allowable world lines obviates timelike curves 
with abrupt breaks where a line going from the past to the future reverses its
direction.
Such world lines would correspond to processes of creation or
annihilation of electron-positron pairs. 
World lines of this shape (``seagull'' configurations) are forbidden in 
classical theory since the principle of last action does not apply to them.}, 
the vacuum polarization responsible for the renormalization of the coupling
constant $e$ is lacking. 
The problem is therefore
reduces to the absorption of the self-energy divergences.

We now verify that $\mu$ and $\delta m$ have identical dimensions. 
The action is dimensionless in units ${\hbar}=1,\,c=1$. 
For the first term of (\ref{W-em}), $[\mu]\,l\,[v]=1$, and, with $[v]=1$, 
we have $[\mu]=l^{-1}$. 
For the third term, $l^{\hskip0.3mm D+1} [A^2]\,l^{-2}=1$, and hence 
$[A]=l^{\hskip0.3mm(1-D)/2}$. 
For the second term, $[e]\,l\,[A]=1$, that is, $[e]=l^{\hskip0.3mm (D-3)/2}$. 
In view of (\ref{delta m}) and (\ref{varphi}), 
$[\delta m]=[e^2]l^{\hskip0.3mm 2-D}=l^{-1}$. 
Thus the singularity of $\delta m(\epsilon)$ can be
cancelled by the singularity of $\mu(\epsilon)$, and (\ref{ren-m}) becomes 
finite.

All troubles with divergences are then over for $D=3$,
and a consistent theory results from the action (\ref{W-em}). 
However, the situation is more intricate for $D=5$. 
For {\it arbitrarily moving} charged particle, the self-energy involves two 
divergent terms. 
The leading divergence is cubic. 
It occurs even in the static case, and is renormalized by $\mu$.
However, there is one further, {\it linear}, divergence \cite{k9}. 
It cannot be removed by the renormalization because the action (\ref{W-em}) 
contains no term with a parameter $\nu$ of the appropriate dimension 
$[\nu]=[e^2]\,l^{-1}=l$. 
The corresponding contribution to the
electromagnetic field momentum $P_\mu$ is proportional to 
\[
{e^2}\,{\epsilon}^{-1}\,(2{\dot a}_\mu+3a^2 v_\mu). 
\]
When compared with (\ref{rig-momentum}), it becomes apparent that
acceleration-dependent Lagran\-gi\-ans involve the parameter $\nu$ enabling
the absorption of the linear divergence. 
Thus a consistent classical $D+1$-dimensional electrodynamics 
for $D+1>4$ can be derived from the action with higher derivatives.

However, what has $D+1>4$ to do with us till we are in four dimensions 
and cannot escape to realms of higher dimensions?
This raises the counter-question: 
Why do we think of our realm four-dimensional?
Whether may four dimensions be illusory? 
However, given $D+1=4$ as a plausible hypothesis, the question
immediately arises: How stands out the case $D=3$ against another dimensions
physically? 
Ehrenfest \cite{Ehrenfest17} was the first to set and try to
solve it. 
The essence of his solution is that. 
No stable composite
particle system can exist in realms with $D>3$, for example, a system
similar to the hydrogen atom: It is imperative that the electron falls to
the nucleus in it.

Greatly simplifying matters, we have to do with the solution of the
relativistic Kepler problem. This is a two-particle problem which can be
reduced to the problem of a single particle moving in the field of the
potential $U(r)$ and specified by the Hamiltonian 
(see, e. g., \cite{Landau}, Sec.\ 39) 
\begin{equation}
H=\sqrt{m^2+\frac{p_\phi^2}{r^2}+p_r^2}+ U(r),  \label{Hamiltonian}
\end{equation}
where ${p_{\phi }}$ and ${p_{r}}$ are the momenta canonically conjugate to
the polar coordinates $\phi $ and $r$. 
Note that ${p_{\phi }}$ is a conserved quantity, the orbital momentum $J$. 
Switching off the dynamics, i.\ e., taking ${p_{r}}=0$ in (\ref{Hamiltonian}), 
we obtain the effective
potential ${\cal U}(r)$ which is convenient for analyzing the particle
behavior near the origin 
\begin{equation}
{\cal U}(r)=\sqrt{m^2+\frac{J^2}{r^2}}+ U(r).  \label{effective}
\end{equation}

There are three alternatives. First, the attractive potential $U(r)$ is more
singular at the origin than the centrifugal term $J/r$. The particle can in
principle orbit in a circle of the radius corresponding to 
${\cal U}_{\hskip0.5mm0}$, the local maximum of the potential 
${\cal U}(r)$. But this
orbiting is unstable, and the fall to the center is highly probable. 
If $E>{\cal U}_{\hskip0.5mm0}$, the fall to the centre is unavoidable.

Second, $U(r)$ is less singular than $J/r$. In particular, for $U(r)=-Ze^2/r$
this means that $Ze^2<J$. The particle executes a stable finite motion. The
fall to the centre is impossible, except when $J=0$.

Third, the singularities of $U(r)$ and $J/r$ are identical, i.\ e., 
$U(r)=-Ze^2/r$, $Ze^2=J$. The particle travels in a stable orbit that passes
through the center.

The quantum-mechanical analysis essentially confirms these conclusions. 
It follows from the solutions of the Schr\"odinger equation 
\cite{GurevichMostepanenko} and relativistic wave equations for particles with
spins 0 and 1/2 \cite{case} that, in the case of sufficiently singular
potentials $U(r)$, bound states form a discrete spectrum extending from $E=m$
to $E=-\infty$. 
The system tends to more advantageous states associated with
successively lower energy levels. As this take place, the dispersion of the
wave function tends to zero as $E_n\to-\infty$. The process resembles the
fall to the center in its classical interpretation.

If the potential $U(r)$ is less singular than the quantum-mechanical
centrifugal term, the spectrum is bounded below. The only distinctive feature
of the quantum-mechanical situation is that there exists a stable ground
state with $J=0$. However, this does not entail the fall to the centre, as
the wave function behaves smoothly in the vicinity of the origin; there is
balance between attraction and zero-point motion.

Since $U(r)=e\,\varphi(r)$ where $\varphi(r)$ is the solution of the 
$D$-dimensional Poisson equation (\ref{varphi}), Ehrenfest inferred from this
that the fall to the center is prevented for $D=3$, but 
the fall is unavoidable for $D>4$. The point $D=3$ is 
critical,
separating realms where stable bound states are feasible from those where
such states are impossible.

The reason for prevention of the fall to the center is that the centrifugal
manifes\-ta\-ti\-ons of kinetic energy (the term $J/r$ or zero-motions)
dominate over attractive forces. 
However, the Hamiltonian (\ref{Hamiltonian}) is essential for such a 
conclusion. 
It is derived from the action (\ref{W-em}). 
But this action is {\it unsuited} for a consistent description of 
electromagnetic interactions for $D>3$. 
This action should be supplemented by terms with higher derivatives, for 
example, terms dependent of the world line curvatures are requisite for $D=5$, 
terms dependent of curvatures and torsions are necessary for $D=7$, etc. 
In the rigid dynamics, the two-particle problem is no longer
Keplerian, it cannot be reduced to the problem of a single particle
orbiting across a plane around the center of mass. 
The problem of two rigid particles in an exact setting is not solved. 
We can made only plausible conjecture of the behavior of such a system. 
The quantity responsible for the centrifugal effect is likely to be more 
singular than $J/r$.
For an acceleration-dependent Lagrangian, it is estimated to be 
$\sim 1/r^3$ \cite{k96}, and hence  the fall to the center can
be prevented for $D=5$ as well.

The presented reasonings are abundantly supplied with simplifications. For
example, when disregarding relativistic effects of the retardation and
radiation and restricting ourselves to the potential picture, we miss the
possibility of the fall to the center due to the dissipation of the particle
energy (recall that the leading impetus to the invention of Bohr's
quantization rules was the problem of the fall of the radiating electron to
the nucleus in the Rutherford model). However, a
more complete analysis would take us away from the major theme. For a more
full discussion of the suppressibility of collapse see \cite{k01}.

A characteristic feature of the rigid dynamics is the Zitterbewegung. 
If the Zitter\-be\-we\-gung occurs in $d$ dimensions, these dimensions may be
considered to be frozen for rectilinear Galilean propagations. 
The last are feasible only in the remaining $D-d$ dimensions. 
It can be shown \cite{KosyakovNesterenko} that the Zitterbewegung of a rigid 
particle with an
acceleration-dependent Lagrangian is possible in two dimensions, and cannot
stretch over larger number of dimensions. 
If we are dealing with $D+1=6$ where every point object executes a 
two-dimensional Zitterbewegung, these two dimensions are effectively
compactified, and, from the point of view of a
center-of-mass observer, the realm is four-dimensional.
However, this effective compactification cannot be realized
on the classical level. 
The reason is that accelerated motions of a charged particle in $D+1=6$ 
is attended with radiation \cite{k9}. 
The particle lost energy, thereby the amplitude of the Zitterbewegung
is diminished, and the motion goes asymptotically to the Galilean regime.

The above construction of a consistent
electrodynamics can with minor
reservations be extended to the classical Yang--Mills theory: It is
necessary to augment the mechanical part of the action of this theory by the
addition of terms with higher derivatives \cite{k9}.

The moral is that turning to the problem of four dimensions of our
world inevitably leads to concepts of the rigid dynamics.

\section{Dressed particles}
\label
{dressed} 
The reader must be familiar with the notion of ``dressed particles'', 
though, most probably, not from classical, but from quantum
electrodynamics, where perturbation series of Feynman diagrams suggests
the view of the electron wrapped up in the coat of electron-positron pairs.
Owing to this coat, the renormalized mass $m$ and charge $e$ of the electron
differ from the corresponding bare quantities $m_{0}$ and $e_{0}$ appearing
in the initial Lagrangian. One may be under the impression that the notion
of the ``dressed particle'' is essentially quantum, since processes of
creation and annihilation of particles occur only in quantum picture.
However, this impression is wrong.

The renormalization of mass takes place in any system with infinite degrees 
of freedom. 
For example, it has
since midnineteenth century been known that a spherical body of mass $m_{0}$
moving with velocity ${\bf v}$ through an ideal fluid behaves as an object
with kinetic energy $(1/2)\,m{\bf v}^{2}$ where $m=m_{0}+\delta m$, that is,
its mass turns out to be augmented by the so called apparent additional mass 
$\delta m$ equal to half the mass of the fluid displaced by the body.
Dynamically, the dragged fluid train is integral part of this
aggregate. 
The quantity $m$ serves as a measure of its inertia, and
the ``bare'' mass $m_{0}$ no longer reveals itself.

The notion of the ``dressed particle'' is equally useful in the classical
field theory with point-like sources. 
As is well known, historically, the
idea of the electromagnetic mass precedes the quantum mechanics, originating
from works by J.\ J.\ Thomson, who based himself on the analogy between the
hydrodynamic medium and the aether (for more detail see, e.\ g., 
\cite{Dresden}). 
We turn to two comparatively simple models of classical
dressed particles to proceed with the discussion of the problem of mass.

\subsection{Dressed charged particle}
\label
{em-dressed} 
The Maxwell--Lorentz theory of $N$ point-like charged
particles is described by the action (\ref{W-em}) with $D=4$. How does the
``dressed particles'' come there? Consider the source (\ref{j-mu}) composed
of a single term. The generic solution to Maxwell's equations (\ref{maxw})
may be represented as $F=F_{ret}+F_{ex}$ where $F_{ret}$ is the retarded
electromagnetic field generated by the source, and $F_{ex}$ is an external
field governed by the free Maxwell equations. The field $F$ should be
regularized (that is, the singularity of the function $F_{ret}$ is smeared
out in some relativistically invariant fashion) and inserted into the
equation obtained by the variation of the action (\ref{W-em}) w.\ r.\ t.\
$x^\mu(s)$, 
\begin{equation}
\mu a^\lambda=e\,F^{\lambda\mu} v_\mu,  
\label{eq-motion-Lorentz}
\end{equation}
to yield, upon the renormalization of mass, Eq.\ (\ref{ren-m}), 
the Abraham--Lorentz--Dirac equation (see, e.\ g., 
\cite{Teitelboim}--\cite{Barut74}): 
\begin{equation}
ma^\lambda-{\frac{2}{3}}\,e^2\bigl({\dot a}^\lambda +v^\lambda a^2 \bigr) 
=e\,F^{\lambda\mu}_{ex} v_\mu.  \label{LD}
\end{equation}

Naively, one believes that Eq.\ (\ref{LD}) describes, as before, the
evolution of mechanical degrees of freedom appearing in the action 
(\ref{W-em}), but takes into account the individual actions on the particle 
of the external and self fields, $F_{ex}^{\mu \nu }$ and $F_{ret}$. 
The role of a finite ``self-interaction'' (or ``back reaction'', or 
``radiation reaction force'', or ``radiation damping force'', etc.) 
\cite{Landau}--\cite{Barut} is attributed to the higher derivative term in 
Eq.\ (\ref{LD}).
Strange as it may seem, this interpretation exists happily for a good century,
despite the fact that it is inconsistent and opens on numerous puzzles and 
paradoxes.

The self-interaction is, by definition, inherent in {\it composite} systems
with reasonable autonomous constituents affecting each other. Such systems
should possess sufficiently great number of degrees of freedom, at least $%
\geq 6$. As to Eq.\ (\ref{LD}), it is an ordinary differential equation
describing the evolution of an object with the number of degrees of freedom
certainly less than 6.

What is the object we are dealing with in actuality? Clearly, it is a
synthetic object because it is characterized by the quantity $m$ involving
the mechanical $\mu $ and field $\delta m$ contributions. This object
originates from the rearrangement of initial degrees of freedom in the
action (\ref{W-em}). It is natural to refer to it as a {\it dressed}
particle. The dressed particle is a stretched object. It can be imagined as
something like the de Broglie ``pilot-wave'' formed by the field train with
a singularity at the point of the charge localization. Dynamical states of
the dressed particle are specified by the four-coordinate of the singularity 
$x^{\mu }$ and the attached to this point four-momentum 
\begin{equation}
p^{\mu }=m\,v^{\mu }-{{\frac{2}{3}}}\,e^{2}a^{\mu }.  \label{p-mu-ren-em}
\end{equation}
The motion of the singularity is described by the equation 
\begin{equation}
\stackrel{\scriptstyle v}{\bot }({\dot{p}}-f)=0  \label{Eq-Newton}
\end{equation}
where $f^{\mu }$ is an external four-force applied to the point $x^{\mu }$.
Indeed, the substitution of (\ref{p-mu-ren-em}) in (\ref{Eq-Newton}) results
in the Abraham--Lorentz--Dirac equation (\ref{LD}) with  $f^{\mu
}=e\,F_{ex}^{\mu \nu }v_{\nu }$. On the other hand, Eq.\ (\ref{Eq-Newton})
is nothing but Newton's second law in the invariant geometric
representation. Equation (\ref{Eq-Newton}) involves only the external force $%
f^{\mu }$ but is deprived of \ an explicit ``self-interaction'' term
occurrence. The dressed particle does not act on itself, it behaves as an 
{\it elementary} entity.

One further reason, advanced by C.\ Teitelboim \cite{Teitelboim}, for the
object with the four-momentum $p^{\mu }$ of the form (\ref{p-mu-ren-em}) to
be singled out in its own right is that the Abraham--Lorentz--Dirac equation
(\ref{LD}) stems from the energy-momentum balance at every point of the
world line: 
\begin{equation}
{\dot{p}}^{\mu }+{\dot{{\cal P}}}_{{}}^{\mu }+{\dot{P}}_{{\rm ex}}^{\mu }=0
\label{loc-mom-balance}
\end{equation}
where the four-momentum of the dressed particle $p^{\mu }$ is defined by (%
\ref{p-mu-ren-em}), the four-momentum of the radiation ${\cal P}_{{}}^{\mu}$
is derived from the Larmor formula 
\begin{equation}
{\cal P}_{{}}^{\mu }=-{{\frac{2}{3}}}\,e^{2}\int_{-\infty }^{s}d\tau
\,v^{\mu }a^{2},  \label{Larmor}
\end{equation}
and the four-momentum $P_{{\rm ex}}^{\mu }$ relates to the integral of the
external Lorentz four-force 
\begin{equation}
{P}_{{\rm ex}}^{\mu }=-\int_{-\infty }^{s}d\tau \,f^{\mu }.  \label{Lorentz}
\end{equation}
Equation (\ref{loc-mom-balance}) reads: The four-momentum extracted from the
external field $-f^{\mu }ds$ is spent on the variation of the four-momentum
of the dressed particle $dp^{\mu }$ and the four-momentum ${\dot{{\cal P}}}%
_{{}}^{\mu }ds$ carried away by the radiation.

The dressed particle can behave in a non-Galilean manner. With $f^\mu=0$,
Eq.\ (\ref{LD}) is satisfied by 
\begin{equation}
v^\mu( s)=\alpha^\mu{\cosh}(w_0\tau_0 e^{s/\tau_0})+ \beta^\mu{\sinh}%
(w_0\tau_0 e^{s/\tau_0})  \label{run-away}
\end{equation}
where $\alpha^\mu$ and $\beta^\mu$ are constant four-vectors that meet the
conditions 
\begin{equation}
\alpha\cdot\beta=0,\quad \alpha^2=-\beta^2=1,  \label{alpha-beta-cond}
\end{equation}
$w_0$ is an initial acceleration magnitude, $\tau_0={2e^2}/{3m}$. The
solution (\ref{run-away})-(\ref{alpha-beta-cond}) describes a runaway
motion, which degenerates to the Galilean regime when $w_0=0$.

It is often asserted that the solution (\ref{run-away}) is ``unphysical'',
because it seems to contradict the energy conservation law: In the absence
of external forces, the particle takes a run with the exponentially growing
acceleration and radiates, that is, the energy of both the particle and
electromagnetic field increases for no apparent reason. Using this line of
reasoning, one keeps in mind either explicitly or implicitly that the
mechanical object possesses the four-momentum $p^{\mu }=mv^{\mu }$, with its
time component ${\cal E}=m\gamma $ being a positive definite quantity.
However, it is beyond reason to insist on the existence of the object with
such a four-momentum. A careful analysis with the use of different
regularization procedures compatible with symmetries incorporated in the
action (\ref{W-em}) leads \cite{Teitelboim} to the selection of an object
possessing the four-momentum of the form (\ref{p-mu-ren-em}) together with
the balance equation (\ref{loc-mom-balance}). As shows this equation, there
is no contradiction with the energy conservation law: The variation of
energy of the dressed particle $dp^{0}$ is equal to the energy carried away
by the radiation $-{\dot{{\cal P}}}_{{}}^{0}ds$. A subtlety is that the
object is characterized by the energy 
\begin{equation}
p^{0}=m\gamma \,\bigl(1-\tau _{0}\,\gamma ^{3}\,{\bf a}\cdot {\bf v}\bigr)
\label{p-0-em}
\end{equation}
which is {\it not a positive definite quantity} (this is scarcely
surprising, if we recall the synthetic origin of the dressed particle). The
indefiniteness of the expression (\ref{p-0-em}) means that the increase of
velocity may occasionally be accompanied by the decrease of energy. It
would, therefore, make no sense to inquire: Where does the particle extract
energy from to accelerate itself? The energy of the self-accelerated dressed
particle is actually {\it diminished}.

Why did this problem not arise for the free gyroscope and rigid particle? As
is shown in Sections 3 and 5, for such objects ${p}^{\mu }=$ const, and the
invariability of $p^{\mu }$ is due to a general reason, the translational
invariance. Since the dependence of $p^{\mu }$ on kinematical variables is
intricate \lbrack see Eqs.\ (\ref{eq-motion}) and (\ref{rig-momentum}%
)\rbrack , the variation of velocity can be compensated by the variation of
higher derivatives in such a way as to respect the condition $p^{\mu }=$
const. As to the dressed particle, in the absence of external forces, its
four-momentum, in general, need not conserve. Now the constant of motion
associated with the translational invariance is, by (\ref{loc-mom-balance}),
the quantity $p^{\mu }+{\cal P}_{{}}^{\mu }$. To illustrate, for the motion
in the regime (\ref{run-away}), the quantity ${\cal P}_{{}}^{0}$ increases
while the quantity $p^{0}$ decreases with the same rate.

When the result of the renormalization of mass, Eq.\ (\ref{ren-m}), is $m=0$%
, the first term in Eq.\ (\ref{LD}) disappears, and, with $f^{\mu }=0$, it
reduces to 
\[
(\stackrel{\scriptstyle v}{\bot}{\dot{a}})^{\hskip0.3mm\mu }=0 
\]
which is the equation of a relativistic uniformly accelerated motion \cite
{Rohrlich}. The world line of the dressed particle with $m=0$ in the absence
of external forces is a hyperbola 
\begin{equation}
v^{\mu}(s)={\alpha^{\mu}}\,{\cosh}\,w_{0}s+{\beta^{\mu}}\,{\sinh}\,w_{0}s,
\quad \alpha\cdot\beta=0,\quad \alpha^{2}=-\beta^{2}=1.
\label{hyperb-motion}
\end{equation}
The curvature $k=w_{0}$ = const of such a world line may be arbitrary. The
radiation goes with a constant intensity determined by the acceleration
squared $a^{2}=-w_{0}^{2}$. As regards the energy of the dressed particle $%
p^{0}$, in view of (\ref{p-0-em}), it is positive in the region of
deceleration, $s<0$, and negative in the region of acceleration, $s>0$.

The reader can verify by a direct calculation that, when moving along the
world lines (\ref{run-away}) or (\ref{hyperb-motion}) for a finite period of
time $\Delta s$, the increase of the radiation energy $\Delta{\cal P}^{0}$
is exactly as the decrease of the energy of the dressed particle $\Delta
p^{0}$.

From (\ref{p-mu-ren-em}) follows that the invariant $v\cdot p$ is a
conserved quantity both in the absence and in the presence of interactions,
because the renormalized mass $m$ is taken to be constant. By contrast, $M=%
\sqrt{p^{2}}$ depends on the form of the world line, that is, it is not
conserved quantity, 
\begin{equation}
M^{2}=m^{2}\,\bigl({1+\tau _{0}^{2}\,a^{2}}\bigr).  \label{M-m}
\end{equation}
(It is remarkable that, for a dressed {\it rigid} particle, $v\cdot p$, $p^2$%
, and any other invariant constructed from $p^\mu$ and kinematical variables 
$v^\mu$, $a^\mu$, etc., are not constants of motion \cite{k9, Nesterenko91},
as distinct from the Abraham--Lorentz--Dirac particle which, fortunately,
does have a conserved invariant quantity $m=v\cdot p$. The problem of inert
properties of a dressed particle in the general case is seen to be quite
nontrivial.)

The expression (\ref{M-m}) shows that, if $\tau _{0}^{2}\,a^{2}<-1$, the
dressed particle turns to a tachyon state with $M^{2}<0$. Note, however,
that the term $\tau _{0}^{2}\,a^{2}$ is very small in the area of
application of classical description. Consider, for example, the Coulomb
interaction of two electrons separated by a distance of order of Compton's
wave length of the electron $1/m$, the minimal allowable in the classical
context separation. We then have the estimate 
\[
\tau _{0}^{2}\,|\,a^{2}|\sim e^{8}\sim 10^{-8}. 
\]
It is clear that the critical acceleration $|a|=1/\tau_{0}$ is inaccessible
here. If it is granted that the class of acceptable world lines is comprised
of smooth timelike curves with the curvature $k$ less than $\tau_{0}^{-1}$,
the solution (\ref{run-away}) falls outside the scope of this class, and the
momentum space of the dressed particle contains no tachyon states, viz.,
states with $p^{2}<0$.

\subsection{Dressed colored particle}
\label
{YM-dressed} 
The action of the $SU({\cal N})$ Yang--Mills--Wong theory
of ${N}$ colored particles is \cite{Balachandran, k8}: 
\begin{equation}
{\cal A}=-\sum_{i=1}^{N}\int\! d s_i\,\biggl(\mu_i\sqrt{v_i\cdot v_i}+ {\rm %
tr}(Z_i\xi^{-1}_i D_{ s}{\xi}_i)\biggr)-{\frac{1}{16\pi}}\int\! d^4x\,{\rm tr%
}(F_{\mu\nu}F^{\mu\nu})  \label{W-YMW}
\end{equation}
where $\xi_i=\xi_i( s_i)$ are time-dependent elements of the gauge group, 
$Z_i=e_i^at_a$,\, $e_i^a$ are those constants whereby the colored charges of
the particles are set $Q_i=\xi_i Z_i\xi_i^{-1}$, and $t_a$ are generators of
the gauge group. The Yang--Mills field strength $F_{\mu\nu}=F_{\mu\nu}^at_a$
is $F_{\mu\nu}= \partial_\mu A_\nu- \partial_\nu A_\mu-ig\,[A_\mu,A_\nu]$
where $g$ is the Yang--Mills coupling constant. The covariant derivative 
$D_{s}$ is given by the formula $D_{s}=d/ds_i+v_i^\mu A^a_\mu t_a$. Since 
$\xi_i$ transforms as $\xi_i\to\xi^{\prime}_i=\Omega^{-1}\xi_i$ under local
gauge transformation, the gauge invariance of the action (\ref{W-YMW}) is
evident.

Despite the similarity of the actions (\ref{W-em}) and (\ref{W-YMW}), they
gives rise to quite distinct theories (the linear equations of electromagnetic
field and the nonlinear Yang--Mills equations). 
This distinction reveals itself most sharply when the theories are 
``decoded'', that is, expressed in terms of exact
solutions. 
Electrodynamics contains only two fundamental
confi\-gu\-ra\-ti\-ons, the plane wave and the Coulomb field.
The former is peculiar to the situation without sources, while the latter 
is inherent in the situation with point-like sources. 
The set of extremals of the action
(\ref{W-YMW}) is much richer. 
Omitting the case in which there is no
external sources, $D^{\lambda} F_{\lambda\mu}=0$, (for results of numerous 
investigations see, e.\ g., \cite{Rajaraman} and references therein),
the situation with the source of the form 
\[
j_{\mu}(x)=\sum_{i=1}^{N}\int\!ds_{i}\,Q_{i}(s_{i})\,v_{\mu}^{i}(s_{i})\,
\delta ^{4}\biggl (x-x_{i}(s_{i})\biggr)
\]
differs from the corresponding situation in electrodynamics in that there
exist two classes of solutions \cite{k8} describing the Yang--Mills
backgrounds of two vacuum phases, cold and hot. 
The solutions corresponding
to the hot phase are fields of the Coulomb type constructed on the Cartan
subgroup of the gauge group. 
For such solutions, all commutators disappear,
and we return to the picture resembling that of electro\-dy\-na\-mics. 
Every result obtained in Sec.\ \ref{em-dressed} is reproduced here with minor
change $e^{2}\rightarrow {\rm tr}\,Q^{2}$.

The solutions of the other class corresponding to the cold phase are
non-Abelian. 
These solutions determine not only field configurations, but
also the colored charges of the sources that generate such
configurations
\footnote{We point out that the parameters $e_i$ in (\ref{W-em}) and $e_i^a$ in (\ref
{W-YMW}) are in no way fixed a priori. There is no restrictions in choosing
the number fields of their values. The solutions describing the cold phase
fixes {\it imaginary} values of the colored charges, $e_i^a=2i/g$. In the
hot phase, it is naturally to ascribe arbitrary {\it real} values to the
quantities $e_i^a$ for stability reasons, for more details see \cite{k8}.}.
When on the subject of a colored particle, we call it {\it quark}, and
omit the particle label $i$. 
We now turn to
the cold phase situation. 
The magnitude of the quark color charge
takes a fixed value 
\begin{equation}
\vert{\rm tr}\,Q^2\vert=\frac{4}{g^2}\,(1-\frac{1}{{\cal N}}).  \label{Q-def}
\end{equation}
The equation of motion for a dressed quark in an external Yang--Mills field 
$F^{\mu\nu}$ is \cite{k1,k8} 
\begin{equation}
m\,\bigl[a^\mu+\ell\,\bigl({\dot a}^\mu +v^\mu a^2 \bigr)\bigr]= {\rm tr}(Q
F^{\mu\nu})\,v_\nu,  \label{eq-motion-YM-part}
\end{equation}
where $m$ is the renormalized mass, and 
\begin{equation}
\ell=\frac{2}{3m}\,\vert{\rm tr} Q^2\vert.  \label{tau-0-YM}
\end{equation}

Equation (\ref{eq-motion-YM-part}) can be written in the form of Newton's
second law (\ref{newton-symbl}) in which 
\begin{equation}
p^\mu= m\,(v^\mu+\ell\,a^\mu)  \label{p-mu-YMW}
\end{equation}
is the dressed quark four-momentum.

Equation (\ref{eq-motion-YM-part}) can be represented as the local
energy-momentum balance 
\begin{equation}
{\dot p}^\mu+{\dot{\wp}}^\mu+{\dot P}^\mu_{{\rm ex}}=0
\label{loc-mom-balance-YM}
\end{equation}
where the four-momentum of the dressed quark $p^\mu$ is given by (\ref
{p-mu-YMW}), 
\begin{equation}
{\wp}^\mu=m\,\ell\int_{-\infty}^{ s} d\tau\, v^\mu a^2,  \label{YM-Larmor}
\end{equation}
\begin{equation}
{P}^\mu_{{\rm ex}}=-\int_{-\infty}^{ s} d\tau\, {\rm tr}(Q
F^{\mu\nu})\,v_\nu.  \label{Lorentz-Wong}
\end{equation}

The balance equations (\ref{loc-mom-balance}) and (\ref{loc-mom-balance-YM})
differ only in their second terms. 
Based on the interpretation of ${\cal P}^{\mu}$ as the four-momentum carried 
away by a divergent wave from the source, ${\wp}^{\mu}$ should be taken as 
the four-momentum conveyed by a convergent wave to the source. 
While part of degrees of freedom of 
electromagnetic fields exists in the form of radiation, the pertinent
degrees of freedom of the Yang--Mills field  in the cold phase play the 
role of a ``negative energy radiation''. 
The balance equation (\ref{loc-mom-balance-YM}) reads: The
four-momentum extracted from the external field $-d{P}_{{\rm ex}}^{\mu}$ is
spent on the variation of the four-momentum of the dressed quark $dp^{\mu}$
and the four-momentum ${\dot{\wp}}^{\mu }ds$ carried away by the ``negative
energy radiation''.

Equation (\ref{eq-motion-YM-part}) with zero right hand side has a solution 
\begin{equation}
v^\mu(s)=\alpha^\mu{\cosh}(w_0\ell\,e^{-s/\ell})+ \beta^\mu{\sinh}(w_0\ell\,
e^{-s/\ell}),  \label{self-disseleration}
\end{equation}
$\alpha^\mu$ and $\beta^\mu$ meet the conditions (\ref{alpha-beta-cond}).
The solution (\ref{self-disseleration}) describes a self-decelerating
motion. 
Although the energy of the dressed quark $p^0$ increases, this increase 
exponentially weaken in time. 
As is seen
from (\ref{loc-mom-balance-YM}), the increase of $p^0$ relates to the
conveyance of energy of the Yang--Mills field attributed to the term 
${\dot{\wp}}^0$.

At first sight, the self-deceleration is an innocent
phenomenon, because the motion becomes almost indistinguishable from 
Galilean in the short run. 
However, the presence of self-decelerations actually jeopardizes 
the consistency of the theory. 
Indeed, as we go to the past, the acceleration increases, and
the intensity of the ``negative energy radiation'' grows along with it.
Thus,  the energy of the Yang--Mills field at any {\it finite} instant is
divergent. 
This is clear from substituting the solution (\ref{self-disseleration}) in 
the integral (\ref{YM-Larmor})
\footnote{It is interesting that runaways do not play a similar role in
electrodynamics with the retarded boundary condition. 
They entail no ``infrared'' divergences. Indeed, the insertion of the solution 
(\ref{run-away}) in the integral (\ref{Larmor}) gives a finite result.}.

Such ``infrared'' divergences cannot be removed from the theory by the
renormali\-za\-ti\-on of physical quantities. On may get ride of them only
by a narrowing the class of acceptable world lines.
Then the solution (\ref{self-disseleration}) would correspond to the world 
line that
is ruled out in advance on the general grounds.

On the other hand, from (\ref{p-mu-YMW}) we have 
\begin{equation}
p^{2}=m^{2}\,\bigl({1+\ell ^{2}\,a^{2}}\bigr).  \label{p-sqr-YM}
\end{equation}
Thus a dressed quark can turn to the tachyon state when $|a|>\ell^{-1}$. By
analogy with electrodynamics, we might require that the class of acceptable
world lines be composed of curves with the curvature less than $\ell ^{-1}$. 
Then states with $p^{2}<0$ would be automatically excluded from the
momentum space.

However, we have no longer phenomenological ground for such restrictions on
the curvature. Equations (\ref{M-m}) and (\ref{p-sqr-YM}) differ only in the
change $\tau _{0}\rightarrow\ell$. But this change radically alter the
situation. From (\ref{Q-def}) and (\ref{tau-0-YM}) follows that $\ell $
depends on the coupling constant as $g^{-2}$. If the coupling is strong, i.\
e., $g\sim 1$, $\ell $ is of order of Compton's wave length of the quark $%
\Lambda _{q}=1/m$, and if $g\ll 1$, $\ell $ is even $g^{-2}$ times greater
than $\Lambda _{q}$. As an illustration, let two quarks, interacting through
the Coulomb-like colored force, be separated by a distance $r$. As is easy
to see, the critical acceleration whereby the quarks turn into tachyon
states is attained at the separation $r\approx |{\rm tr}Q^{2}|/m$, which is
more than Compton's wave length of the quark by a factor of $g^{-2}$.
Effects associated with great quark accelerations, the critical value $%
|a|=\ell^{-1}$ included, fall within the area of application of classical
theory.

Thus the conversion quarks to the tachyon state may provide some insight
into subnuclear physics. A plausible assumption is that, crossing the point $%
p^{2}=0$ corresponds to the transition between the cold and hot phases
rather than the would-be conversion of the quark to the tachyon state \cite
{k2001}.

\section{Concluding remarks}
\label
{conclusion} 
We began with the assertion that $M$ alone cannot provide an 
exhaustive account of inert properties of point objects. 
Two invariants, the mass $M$ and the rest mass $m$, played a key role in the 
following discussion. 
For a Galilean particle, $M=m$, both of these quantities being identical to the
operationally well defined Newtonian mass. 
On the other hand, for the Frenkel spinning particle, $m>M$, and, therefore, 
the relation of $m$ and $M$ to experimentally
measured quantities is an open question. 
The situation with rigid particles is somewhat simpler, because, in the 
absence of external forces, the only conserved quantity is $M$. 
In the Lagrangian formalism, we encountered dimensionfull parameters $\mu$ and 
$\nu$, as well as the time-dependent monad $\eta^{-1}$ and
Lagrangian multiplier $\chi $. 
While these and similar quantities are formally related to $m$ and $M$, 
they are of little physical concern. 
The reason for this is clarified by the example of the bare mass $\mu$
which ceases from being a mere number and becomes a
function of a regularization parameter $\mu(\epsilon)$.
The dependence on $\epsilon$ is taken such that adding $\mu(\epsilon)$ to the 
self-energy $\delta m(\epsilon)$ results in the cancellation of their 
singularities rendering the renormalized mass $m$ finite. 
(This seemingly awkward
regularization-renormalization procedure is in fact an integral part of 
local field theories both on classical and quantum levels; albeit,
mathematically, we are dealing with a quite respectable procedure of
extraction of finite values based on the solution of the fundamental
Riemann-Hilbert problem \cite{Connes}.) 
In addition, the relation of $m$ and $M$ to the gravitation mass $M_{{\rm g}}$ 
was twice cursorily touched. 
A significant reduction of the number of quantities characterizing
inert properties of non-Galilean objects can hardly be conceived.

`Gracious me! Why do we go into details of inertia of non-Galilean objects, 
even though
no one {\it observed} Zitterbewegungs, runaways, and other
extavagant regimes of free evolution?' the perplexed reader may interrupt at 
this point.

Surely anyone {\it endeavored} to observe them?

`Yes, nothing has been heard of such experiments,' the sceptical reader may
continue. 
`But, is it really required a particular contrivance? 
Why are such regimes {\it not immediately evident} from fleeting 
glance?'

Surely anyone saw Galilean motions with the naked eye? 
Everyday observations convince us: In the absence of external forces, 
bodies {\it are at rest}. 
One day Aristotle arrived at this conclusion, and then, over 2000 years,
none cast doubts on this subject. 
A lot of the credit must go to Galilei
since he had the courage to make far-reaching {\it extrapolations} from
everyday observations and verify the idea of the uniform motion in
experiments {\it specially adapted} to the clarification of this issue.

Although the Zitterbewegung fails to be visible, there are circumstantial
evidences that such a regime is yet feasible. 
Unfortunately, the frequency peculiar to the Zitterbe\-we\-gung is of order 
of Compton's wave length of the object. 
This casts suspicion on the classical interpretation of this phenomenon. 
Nevertheless, the theoretical framework is large enough to expect that 
objects executing a Zitterbewegung with certainly classical value of 
frequency do exist.

As to processes with growing accelerations, they are inherent in unstable
systems, specifically systems with two phases which are capable of a phase 
transition  (e. g., the early Universe inflation \cite{Guth, Linde}, 
deconfinement \cite{Meyerortmanns}, etc.). 
It is not unlikely that such phenomena might be conveniently
expressed in terms of self-accelerated dressed particles. 
Moreover, cosmological objects are every bit well suited for the role of
self-accelerated particles. 
Indeed, great efforts are made to
explain the recent discovery of the {\it accelerated} expansion of the Universe
in models with the $\Lambda $-term \cite{Straumann}. 
An alternative explanation may be quite simple: Cosmological objects execute
self-accelerated motions, analogous to the runaways of dressed charged
particles, Eqs.\ (\ref{run-away}) and (\ref{hyperb-motion}). 
Notice, we are
dealing with objects (supernovas, galaxies, quasars, etc.)
possessing internal angular momenta, thus their non-Galilean regimes may intricately
combine the Zitter\-be\-we\-gung and motion with increasing velocity.

`And yet the condition of {\it classicality} is essential for this discussion 
altogether. 
However, fundamental laws of the Nature are quantum. 
The four-momentum $p^{\mu }$ is the only
well-defined dynamical variable in quantum theory, hence only $M$ is
relevant here (the four-velocity $v^{\mu }$ is not a well-defined quantum
variable). 
Thus the problem $M\neq m$ is far-fetched. 
Although the notion of the non-Galilean particle exists 
(albeit under another names)  in theoretical physics already over some 75 
years, no  tangible thing underlies it. 
What is the use of it? 
Should we ever trouble with the archaisms
like the Abraham--Lorentz--Dirac equation, or Frenkel's particle? 
Maybe, it
is appropriate time to get ride of this theoretical rubbish,' the
irrepressible reader casts his further doubt.

It might be well to recall at this point that there are three radically
different views of the nature of our world. 
One of them asserts that the most profound grasp of the physical reality is 
ensured by classical, deterministic laws. 
They form the fundamental level of cognition. 
One should establish the so called ``hidden variable'' theory to describe it. 
Quantum theory has a phenomenological status, it must be found by averaging 
over the hidden variables. 
In the late 1950s, this viewpoint was vigorously advocated and
elaborated by L.\ de Broglie, J.\ Vigier \cite{Broglie}, and especially D.\
Bohm \cite{Bohm}. 
In modern times it was revived by G.\ 't Hooft \cite{Hooft99}, who maintains 
that deterministic, not quantum, states are the primary states in the 
sub-Planckian domain (with sizes $l<l_{{\rm P}}=1.6\times 10^{-33}$ cm).

The opposite view is that our world is quantum. Macroscopic objects appear
to be classical only effectively. Such classical manifestations are
explained by the so called decoherence \cite{Zurek, Guilini}. This view is
presently very popular. It is supported by results of experimental tests of
the Bell inequalities \cite{Bell, Clauser} (which, admittedly, do not lower
the enthusiasm of adherents of the deterministic viewpoint, see, e.\ g.,
counter-arguments by 't Hooft \cite{Hooft99}). However, this view is
difficult to accept when the human being or the Universe are concerned. With
all the willingness to fall a victim to science, the present author would
not dare to subscribe to the paper as ``a superposition of alive and dead
Kosyakov''. And you, the reader, are you really inspired with the role of a
``decohered'' {\it homo sapiens}?

At last, the third paradigm is based on the {\it coexistence} of the
classical and quantum ontologies. In other words, we are dealing with two
realms. In the classical realm, everything happens unambiguously, at least a
given object certainly exists at the given place and at the given instant,
and its individuality is preserved. In the quantum realm, every process (the
being of objects included) is characterized by some probability amplitude.
The individuality of a quantum object, say, a given electron, is not ensured
since it is identical to any one of real or virtual electrons (that is,
electrons that might not exist at the given instant certainly, but is ready
to appear due to the electron-positron pair creation, muon decay, etc.).
This paradigm is due to founders of the ``Copenhagen interpretation'' who
repeatedly argued for the treatment of quantum objects on equal terms with
macroscopic  classical devices. A link between the classical
and quantum realms is offered by the so called holographic principle.
According to this principle ('t Hooft \cite{Hooft} and L.\ Susskind \cite
{Susskind} were the fist to enunciate it in the context of quantum gravity),
the information on what takes place inside some volume can be projected onto
the boundary, and we have actually to do with the projection of a 
{\it classical} picture in the bulk onto a {\it quantum} picture in the surface 
\cite{k0}. 
The same physical reality may appear either classical or quantum, 
being imbedded in spacetimes of, respectively, $D+1$
and $D$ dimensions.

Thus, to declare the supremacy of quantum notions over classical ones is to
discrimi\-na\-te against the ``Copenhagen'' and ``deterministic'' minorities
which involve not only scientific marginals.

One further comment on the troubles with old-fashioned concepts which are
still not incarnated in observable objects is in order. It is interesting
that the physical community fairly rich in lovers of these theoretical
``relics''. The magnetic monopole was invented by Dirac 70 years ago, the 't
Hooft--Polyakov monopole is already over 25, and, although the experiment
make no hint about the existence of these objects, whether have we not
enough people who argue about the monopole, study its properties, and
suggest resolutions of numerous problems, from subnuclear to cosmological,
as if we deal with a real particle? However, could anybody bring
himself/herself to call the monopole (together with a number of another
somewhat obsolete things of the high energy physics props, e.~g., Higgs
bosons, axions, super\-sym\-met\-ric particles, etc.) the theoretical
``rubbish''?

The all-powerful vogue can turn  ``the well forgotten out-of-date'' to some
up-to-date. 
Who remembered the Born--Infeld electrodynamics 15 years ago?
One might confuse it with the Mie electrodynamics or, at best, elicit a
vague recollection of something ``maybe non-local, maybe nonlinear, or maybe
gauge non-invariant''. 
However, this seemingly forgotten name
is now flashy again in leading physical journals. 
The point is the Born--Infeld Lagrangian emerges in the low energy limit of the superstring
theory.

Peremptory decisions and ``death sentences'' especially those related to an
authorita\-ti\-ve scientist may be detrimental to his own reputation and
studies of his colleagues. There is a great number of precedents. We turn to
two of them.

One day, a young theorist A.\ Salam came to the formidable W.\ Pauli to
submit to him a daring idea of the two-component neutrino. Pauli responded
with a note urging the visitor to ``think of something better''. Discouraged
Salam delayed his publication, and the credit for discovery of the parity
violation fell to Lee and Yang \cite{Duff}. Recall, this was just Pauli who
derived the equation of the two-component massless spinor field 25 years
prior to this event and who repudiated it at once, taking the violation of
mirror symmetry to be absurdity. Giving up this equation for lost, Pauli
turned down any proposal of its physical application.

In the attempt to build a model of static Universe, Einstein introduced the
cosmological constant ($\Lambda $-term) into the gravitation equations 
in 1917. 
Nobody felt the need or even naturalness for this step at that time. 
This was a likely reason why A.\ Friedmann concentrated
on a nonstationary expanding model of Universe described by a solution to
the gravitation equations with zero $\Lambda $-term. Einstein felt something
``suspicious'' in this solution, and he expressed his feeling in his comment
of Friedmann's paper. 
Later on, Einstein accepted both the very idea of
nonstationary Universe and Friedmann's solution, but went into another
extreme and considered the $\Lambda $-term to be his greatest mistake.
Following Einstein, most of theorists brought hastily the $\Lambda $-term in
the category of regrettable {\it ad hoc} constructions. This state of the
art remained unchanged for about 40 years until Ya.\ B.\ Zel'dovich \cite
{Zel'dovich} observed that allowing for zero oscillatory modes makes the
presence of the $\Lambda $-term in quantum gravity inevitable. Since then,
the accounting for the $\Lambda $-term is a central (and challenging)
problem in quantum gravity and cosmology \cite{Weinberg}--\cite{Chernin}.

\vskip5mm 
\noindent 
{\Large {\bf Acknowledgments}} \vskip2mm 
\noindent 
I am indebted to V.\ G.\ Bagrov, A.\ O.\ Barut, T.\ Goldman, G.\ V.\ Efimov, 
I.\ B.\ Khriplovich, V.\ V.\ Nesterenko, F.\ Rohrlich, R.\ Woodard, and 
H.\ D.\ Zee for discussions of this subject at different time. 
This work is
supported in part by the International Science and Technology Center,
Project {\#} 840.

\end{document}